\documentclass[reprint,nofootinbib,superscriptaddress,amsmath,amssymb,aps,prd]{revtex4-2}
\usepackage{amsmath,mathrsfs,amsbsy,color,graphicx,bm,amsthm,amsfonts}
\usepackage{bbm}
\usepackage{times}
\usepackage{units}
\usepackage{dcolumn}
\usepackage{graphicx}
\usepackage{epsfig}
\usepackage{epstopdf}
\usepackage[colorlinks,linkcolor=red,anchorcolor=green,citecolor=blue,CJKbookmarks=True]{hyperref}
\DeclareMathSymbol{\shortminus}{\mathbin}{AMSa}{"39}
\usepackage{mathrsfs}
\usepackage{braket}
\usepackage{amssymb}
\usepackage{txfonts}
\usepackage{float}
\usepackage{enumitem}
\usepackage{multirow}

\begin{document}
\title{Can the signatures of quantum superposition be detected through correlation harvesting?}

\author{Yu Tang}
\affiliation{Department of Physics, Key Laboratory of Low Dimensional Quantum Structures and Quantum Control of Ministry of Education, Hunan Research Center of the Basic Discipline for Quantum Effects and Quantum Technologies, and Synergetic Innovation Center for Quantum Effects and Applications, Hunan Normal
		University, Changsha, Hunan 410081, P. R. China}

\author{Wentao Liu}
\affiliation{Department of Physics, Key Laboratory of Low Dimensional Quantum Structures and Quantum Control of Ministry of Education, Hunan Research Center of the Basic Discipline for Quantum Effects and Quantum Technologies, and Synergetic Innovation Center for Quantum Effects and Applications, Hunan Normal
		University, Changsha, Hunan 410081, P. R. China}

\author{Zhilong Liu}
\affiliation{Department of Physics, Key Laboratory of Low Dimensional Quantum Structures and Quantum Control of Ministry of Education, Hunan Research Center of the Basic Discipline for Quantum Effects and Quantum Technologies, and Synergetic Innovation Center for Quantum Effects and Applications, Hunan Normal
		University, Changsha, Hunan 410081, P. R. China}
        
\author{Jieci Wang}
\email{jcwang@hunnu.edu.cn (Corresponding author)}
\affiliation{Department of Physics, Key Laboratory of Low Dimensional Quantum Structures and Quantum Control of Ministry of Education, Hunan Research Center of the Basic Discipline for Quantum Effects and Quantum Technologies, and Synergetic Innovation Center for Quantum Effects and Applications, Hunan Normal
		University, Changsha, Hunan 410081, P. R. China}

\begin{abstract}

In this paper, we explore correlation harvesting in quantum superposition, specifically focusing on the entanglement and mutual information extracted by two Unruh-DeWitt detectors interacting with a quantum field in a mass-superposed BTZ black hole spacetime.
Our findings reveal that the superposed nature of spacetime induces constructive interference between the field modes that can significantly enhance the entanglement harvesting relative to a single spacetime background.
In contrast to entanglement, the mutual information obtained in spacetime superposition is influenced by the proper distance between the two detectors.
While the mutual information harvested in a superposed spacetime remains lower than that in a single spacetime when the proper distance between detectors is small, it exceeds that in a single spacetime for specific mass ratios as the distance increases.
Notably, we find that both entanglement and mutual information harvesting reach their maxima when the final spacetime superposition state is conditioned to align with the initial spacetime state.

\end{abstract}
\vspace*{0.5cm}
\maketitle
\section{Introduction}
The Reeh-Schlieder theorem \cite{Reeh:1961ujh} in quantum field theory reveals that the vacuum state of a free quantum field exhibits maximal violation of Bell's inequalities, thereby demonstrating profound nonlocal correlations across spacelike-separated regions.
Subsequently, researchers discovered that vacuum correlations can be transferred to physical systems, which led to the development of correlation harvesting---a protocol that extracts correlations from the quantum scalar vacuum using detectors \cite{Valentini:1991eah,Reznik:2002fz,Reznik:2003mnx,Salton:2014jaa,Pozas-Kerstjens:2015gta}.
The quantum resource harvesting protocol, initially developed using the Unruh-DeWitt (UDW) particle detector model has been successfully extended to curved spacetime scenarios \cite{Henderson:2017yuv}, reinforcing its significance as a key subfield in relativistic quantum information \cite{Fuentes-Schuller:2004iaz,Ahn:2008zf,Zhou:2022nur,Zhang:2020xvo,Gallock-Yoshimura:2021yok,Cong:2018vqx,Svidzinsky:2024tjo,Kukita:2017etu,Xu:2020pbj,Tjoa:2020eqh,Foo:2020dzt,Foo:2020xqn,Zhou:2021nyv,Wu:2022lmc,Wu:2022xwy,Chen:2023xbc,Wu:2023sye,Wu:2023spa,Wu:2024BF,Liu:2024pse,Liu:2024yrf,Tang:2025eew,Li:2025bzd,Wu:2024qhd,Liu:2024wpa}.
Numerous investigations confirm that the harvesting efficiency for vacuum correlations is governed critically by the detector trajectory, its energy gap, and—fundamentally—the geometric structure of spacetime \cite{VerSteeg:2007xs,Henderson:2018lcy,Ng:2018drz,Martin-Martinez:2015qwa,Robbins:2020jca,Henderson:2022oyd,Cong:2020nec,Liu:2022uhf,Maeso-Garcia:2022uzf,Liu:2023zro,Lindel:2023rfi,Ji:2024fcq,Wu:2024whx,Naeem:2022drs,Bueley:2022ple,Liu:2025bpp,Chakraborty:2024fsh}.
This phenomenon enables rigorous investigation of the dependence of quantum nonlocality on global spacetime topology.

However, due to the inherently local nature of general relativity, a complete quantum gravity theory may be required to fully determine the universe’s global topology.
While one approach to quantum gravity involves developing a new gravitational framework, such as string theory and loop quantum gravity \cite{Aharony:1999ti,Berkovits:2000fe,Surya:2019ndm,Lewandowski:2022zce,Zhang:2024khj,Zhang:2024ney,Liu:2024soc,Liu:2024iec}, another focuses on extensive research exploring the phenomenology of quantum gravity from an operational perspective \cite{Henderson:2020zax,Giacomini:2020ahk,Giacomini:2021aof}.
This approach operationally defines observables through measurable quantities specified by theoretical apparatus, such as detectors, rods, and clocks, by exploring quantum-gravitational physics via a ``bottom-up'' approach, such as investigating time quantization using a clock in a superposition of localized momenta \cite{Smith:2019imm}, spacetime metric reconstruction through quantum field correlations \cite{Kempf:2021xlw}, and violations of classical causal order due to superpositions of massive bodies \cite{Zych:2017tau}, favoring gradual progress over a comprehensive top-down theory.

Considering that gravity is fundamentally a theory of spacetime geometry, any forthcoming quantum gravity theory should naturally incorporate the fundamental principles of quantum superposition into the framework of spacetime.
This integration gives rise to ``spacetime superpositions'', wherein distinct spacetime geometries, not related by a global coordinate transformation, are coherently combined in a quantum superposition \cite{Christodoulou:2018cmk,Belenchia:2018szb,Giacomini:2021gei}.
A recent study \cite{Chakraborty:2024fsh} has for the first time investigated entanglement harvesting in superposed Minkowski spacetime, demonstrating that the superposition property of spacetime induces interference effects which can significantly enhance entanglement for the field.
This work provides a more profound understanding of the interconnection between quantum nonlocality and the global structure of spacetime.
In addition, J.Foo { \it et al.} \cite{Foo:2021exb} has constructed the quantum superposition of two nonrotating, chargeless Bañados-Teitelboim-Zanelli (BTZ) black holes with distinct masses, and the response of an UDW detector in this configuration has been subsequently analyzed.
Their results demonstrate that the detector's response manifests discrete resonances at rational ratios of the superimposed masses.
This phenomenon offers a novel perspective for understanding the fundamental nature of quantum gravity.
Inspired by expectations of quantum gravity, we study a related problem: can the signatures of quantum superposition be discerned during the correlation harvesting process?

To address this question, we investigate a pair of decoupled UDW detectors within a background of mass-superposed BTZ black holes and compute entanglement  and mutual information harvested as functions of various physical parameters.
We demonstrate the effects of spacetime superpositions on harvested correlations using numerical calculations.
Building on these studies, we investigate potential signatures of quantum superposition within the framework of correlation harvesting processes.
This paper is organized as follows: 
we first review the theory of mass-superposed BTZ black holes and the construction of automorphic fields on these spacetimes in Sec. \ref{sec2}.
In Sec. \ref{sec3}, we review the model for coupling the UDW detector to a quantum-controlled superposition of spacetimes and apply this to the mass-superposed BTZ spacetime.
In Sec. \ref{sec4}, we then analyze the correlation properties of the final entangled state of the two detectors to determine how the parameters encoding our spacetime superposition affect correlation harvesting.
Finally, we summarize our findings in Sec. \ref{sec5}.
Throughout this article, we utilize natural units, $\hbar=k_{B}=c=G=1$.

\section{QUANTUM FIELDS ON BTZ SPACETIME AND ITS SUPERPOSITION} \label{sec2}
	The BTZ spacetime \cite{Banados:1992wn,Banados:1992gq} is obtained as a quotient of anti-de Sitter-Rindler spacetime under the identification $\Gamma_{M}:\phi\to\phi+2\pi\sqrt{M}$. The metric is
		\begin{equation}
		\mathrm{d}s^2=-f(r)\mathrm{d}t^2+f(r)^{-1}\mathrm{d}r^2+r^2\mathrm{d}\phi^2, \label{metric}
		\end{equation}
	where $f(r)=(r^{2}/l^{2}-M)$, $l$ is the anti-de Sitter (AdS) length scale and $M$ is the mass of BTZ black hole.
	To formulate a quantum field theory in the BTZ spacetime background, we consider an automorphic field $\hat{\phi}^{M_{i}}(\mathbf{x})$, constructed from a massless scalar field $\hat{\psi}$ in (2+1)-dimensional AdS spacetime $(\mathrm{AdS}_3)$ via the identification $\Gamma_{M_{i}}$, yielding \cite{Lifschytz:1993eb}
		\begin{equation}
		\hat{\phi}^{M_i}(\mathbf{x}):=\frac{1}{\sqrt{\mathcal{N}}}\sum_n\eta^n\hat{\psi}(\Gamma_{M_i}^n\mathbf{x}), \label{field}
		\end{equation}
	where $\mathbf{x}=(t,r,\phi)$, $\mathcal{N}=\sum_n\eta^{2n}$ is a normalization factor and $\eta=\pm1$ denotes an untwisted (twisted) field (for simplicity, we
consider only the $\eta=+1$ case here). 
	To obtain the Wightman functions, we have \cite{Lifschytz:1993eb}
		\begin{align}
		W_{\mathrm{BTZ}}^{M_i}(\mathbf{x},\mathbf{x}^{\prime})&=\frac{1}{\mathcal{N}}\sum_{n,m}\eta^n\eta^mW_{\mathrm{AdS_3}}(\Gamma_{M_i}^n\mathbf{x},\Gamma_{M_i}^m\mathbf{x}^\prime)\notag \\&=\frac{1}{\mathcal{N}}\sum_{n,m}\eta^n(\eta^n\eta^m)W_{\mathrm{AdS_3}}(\Gamma_{M_i}^n\mathbf{x},\Gamma_{M_i}^n\Gamma_{M_i}^m\mathbf{x}^{\prime})\notag \\&=\frac{1}{\mathcal{N}}\sum_{n,m}\eta^{2n}\eta^{m}W_{\mathrm{AdS_3}}(\mathbf{x},\Gamma_{M_i}^m\mathbf{x}^{\prime})\notag \\&=\sum_m\eta^mW_{\mathrm{AdS_3}}(\mathbf{x},\Gamma_{M_i}^m\mathbf{x}^{\prime}),\label{Wightman}
		\end{align}
	where $\Gamma_{M_{i}}^{n}:(t,r,\phi)\to(t,r,\phi+2\pi n\sqrt{M_{i}})$ in a BTZ spacetime with black hole mass $M_{i}$.
	
	We investigate the quantization of a field on a background formed by the superposition of BTZ spacetimes characterized by distinct black hole masses.
	The black hole–quantum field system can be described in the tensor product Hilbert space $\mathcal{H}=\mathcal{H}_{\mathrm{BH}}\otimes\mathcal{H}_{F}$, where we consider the black hole to be in a superposition of two mass states $\left|M_{1}\right\rangle$ and $\left|M_{2}\right\rangle$ while the field is in the $\mathrm{AdS_3}$ vacuum $\left|0\right\rangle$. On the joint Hilbert space $\mathcal{H}=\mathcal{H}_{\mathrm{BH}}\otimes\mathcal{H}_{F}$, we need to condition our field operator on the state of the spacetime, yielding the relation
		\begin{equation}
		\hat{\phi}(\mathbf{x})=\sum_{i=1,2}\hat{\phi}^{M_i}(\mathbf{x})\otimes|M_{i}\rangle\langle M_{i}|,\label{fullfield}
		\end{equation}
an analogous procedure as in (\ref{Wightman}) yields \cite{Foo:2021exb,Zhang:2024hkw}

		\begin{align}
		W_{\mathrm{BTZ}}^{M_1M_2}(\mathbf{x},\mathbf{x}^{\prime})&=\frac{1}{\mathcal{N}}\sum_{n,m}\eta^n\eta^mW_{\mathrm{AdS_3}}(\Gamma_{M_1}^n\mathbf{x},\Gamma_{M_2}^m\mathbf{x}^\prime)
		\notag \\&=\frac{1}{\mathcal{N}}\sum_{n,m}\eta^{n}\eta^{m}\frac{1}{4\pi l \sqrt{2}} \Pi(\Gamma_{M_1}^n\mathbf{x},\Gamma_{M_2}^m\mathbf{x}^{\prime}),\label{Wightmanspuer}
		\end{align}
	where 
		\begin{align}
		\Pi(\Gamma_{M_1}^n\mathbf{x},\Gamma_{M_2}^m\mathbf{x}^{\prime})=\frac{1}{\sqrt{\sigma\left(\Gamma_{M_1}^{n}\mathbf{x},\Gamma_{M_2}^{m}\mathbf{x}^{\prime}\right)}}\!-\!\frac{\zeta}{\sqrt{\sigma\left(\Gamma_{M_1}^{n}\mathbf{x},\Gamma_{M_2}^{m}\mathbf{x}^{\prime}\right)+2}},\label{wightman}
		\end{align}
		\begin{align}
		\sigma(\Gamma_{M_1}^n\mathbf{x},\Gamma_{M_2}^m\mathbf{x}^{\prime})=&\frac{r_Ar_B}{\sqrt{M_1}\sqrt{M_2} l^2}\cosh\bigg[2\pi(n\sqrt{M_1}-m\sqrt{M_2}) \notag\\
&+(\phi_A-\phi_B)\bigg]\!-\!1-\sqrt{\tfrac{r_A^2}{M_1 l^2}-1}\sqrt{\tfrac{r_B^2}{M_2 l^2}-1}\notag\\
&\times\cosh\left[\tfrac{\sqrt{M_1}t_A-\sqrt{M_2}t_B}{l}\right].\label{sigma}
		\end{align}
	The parameter $\zeta\in\{-1,0,1\}$ specifies the boundary condition for the field at the spatial infinity: Neumann $(\zeta=-1)$, transparent $(\zeta=0)$, and Dirichlet $(\zeta=1)$.
	Noting the two different isometries, $\Gamma_{M_1}$ and $\Gamma_{M_2}$, corresponding to the superposed masses, $M_{1}$ and $M_{2}$, that act on the coordinates of the field operators.
	This means the mass-superposed black hole, wherein each mass specifies a distinct classical solution to the Einstein field equations, yields an amplitude from the superposition that corresponds to the associated spacetime state \cite{Foo:2021exb}.
	
\section{UNRUH-DEWITT DETECTOR IN SUPERPOSED BTZ SPACETIME} \label{sec3}	
	To couple matter to the quantum black hole-field system, we consider two decoupled UDW detectors \cite{Unruh:1976db,DeWitt:1980hx}, labeled A and B, associated with the detector Hilbert spaces
$\mathcal{H}_{A}$ and $\mathcal{H}_{B}$, respectively.
	The full Hilbert space of the system is given by $\mathcal{H}_{A}\otimes\mathcal{H}_{B}\otimes\mathcal{H}_{F}\otimes\mathcal{H}_{S}$, which is a tensor product of 
the detector, field, and spacetime, respectively.
	The coupling between the spacetime superposition, field, and detector is described by the following interaction Hamiltonian
		\begin{equation}
		\begin{aligned}
		H_{D}^{I}(\tau_{D})=&\lambda\Upsilon_{D}(\tau_{D})(\sigma_{+}(\tau_{D})+\sigma_{-}(\tau_{D}))\\&\otimes\sum_{i=1,2}\hat{\phi}^{M_{i}}(x_{D}(\tau_{D}))\otimes|M_{i}\rangle\langle M_{i}|,\label{int}
		\end{aligned}
		\end{equation}
	where $D=A,B$ and $\sigma_+(\tau_D)=e^{\mathrm{i}\Omega_D\tau_D}|1\rangle\langle0|$, $\sigma_-(\tau_D)=e^{-\mathrm{i}\Omega_D\tau_D}|0\rangle\langle1|$.
	Here, $\lambda$ is a coupling constant, $\tau_{D}$ is the proper time in the detector’s reference frame, and $\Upsilon_{D}(\tau_{D})$ is the time-dependent switching function that mediates the interaction.
	The total interaction Hamiltonian for the two detector-field-spacetime system is given by \cite{Martin-Martinez:2018gzb}
		\begin{equation}
		H_{\mathrm{tot}}^I(t)=\frac{d\tau_\mathrm{A}}{dt}{H}_\mathrm{A}^{I}\left[\tau_\mathrm{A}(t)\right]\otimes \mathbf{I}_{B}+\mathbf{I}_{A}\otimes \frac{d\tau_\mathrm{B}}{dt}{H}_\mathrm{B}^{I}\left[\tau_\mathrm{B}(t)\right],\label{Hami}
		\end{equation}
	where $\mathbf{I}_{A}$ and $\mathbf{I}_{B}$ is the identity operators acting on the Hilbert spaces $\mathcal{H}_{A}$ and $\mathcal{H}_{B}$, respectively.
	The time evolution is given by the unitary
		\begin{align}
		U=&\hat{\mathcal{T}}\bigg[\exp\bigg\{-i\int dt\bigg(\frac{d\tau_A}{dt}\bigg)H_A^{I}(\tau_A(t))\otimes \mathbf{I}_{B}\notag
		\\&+\mathbf{I}_{A}\otimes\left(\frac{d\tau_B}{dt}\right)H_B^{I}(\tau_B(t))\biggr\}\biggr],\label{U}
		\end{align}
	where $\hat{\mathcal{T}}$ is a time-ordering operator.
	Given that the coupling strength $\lambda$ is small, we can expand the time evolution operator $U$ using a Dyson series \cite{Martin-Martinez:2020pss},
		\begin{equation}
		U=\mathbf{I}-i\int_{\mathbb{R}}dtH_{\mathrm{tot}}^I(t)-\int_{\mathbb{R}}dt\int_{-\infty}^{t}dt^{\prime}H_{\mathrm{tot}}^I(t)H_{\mathrm{tot}}^I(t^{\prime})+O(\lambda^{3}).\label{U2}
		\end{equation}
	Then, we assume that both the detector and the scalar field system are in their respective ground states, giving the total system’s initial state as
		\begin{equation}
		|\Phi_i\rangle=|0\rangle_A\otimes|0\rangle_B\otimes|0\rangle_F\otimes|s_i\rangle,\label{istate}
		\end{equation}
	where
		\begin{equation}
		|s_i\rangle=\cos\theta|M_1\rangle+\sin\theta|M_2\rangle,\label{spacetime}
		\end{equation}
	is an arbitrary superposition of the two spacetime states.
	To streamline calculations and maintain a manageable parameter space, we have opted to neglect the complex phase, thus avoiding increased computational complexity while preserving the essential physics of the superposed transition probability \cite{Chakraborty:2024fsh}.
	
	After time evolution of (\ref{istate}) under (\ref{U2}), the final state is given by
		\begin{equation}
		|\Phi_f\rangle=\sum_nU^{(n)}|\Phi_i\rangle=\sum_n\lambda^n|\Phi_f^{(n)}\rangle,\label{fstate}
		\end{equation}
	where the $U^{(n)}$s are the terms of order $\lambda^n$ in the unitary operator $U$.
	To obtain the reduced density matrix for the joint detector state, we trace out the field degrees of freedom while conditioning on the control degree of freedom associated with spacetime states, i.e.,
		\begin{equation}
		|s_f\rangle=\cos\varphi|M_1\rangle+\sin\varphi|M_2\rangle.\label{fspacetime}
		\end{equation}
	The joint state of the detectors $\rho_{AB}=\mathrm{Tr}_{\phi}[\langle s_{f}|U|\Phi_{i}\rangle\times\langle\Phi_{i}|U^{\dagger}|s_{f}\rangle]$ written in the basis $|0,0\rangle,|0,1\rangle,|1,0\rangle,|1,1\rangle,$ and up to $\mathcal{O}(\lambda^{2})$ (Appendix \ref{appa1}) is given by
		\begin{equation}
		\rho_{AB}=
		\begin{pmatrix}
		P_G&0&0&\mathcal{M}\\0&P_B&\mathcal{L}_{\mathrm{AB}}&0\\0&\mathcal{L}_{\mathrm{AB}}^*&P_A&0\\\mathcal{M}^*&0&0&0
		\end{pmatrix}+\mathcal{O}(\lambda^{4}),\label{matrix}
		\end{equation}
	where the respective terms are given by
		\begin{align}
		P_G&=(a+b)^2-2\lambda^2\left[(a^2+ab)\sum_DP_D^{M_1}+(b^2+ab)\sum_DP_D^{M_2}\right],\label{pg}\\
		P_D&=\lambda^2(a^2P_D^{M_1}+b^2P_D^{M_2}+2abP_D^{M_1M_2}),\label{pd}
		\end{align}
		\begin{align}
		\mathcal{L}_{\mathrm{AB}}=&\lambda^{2}\int dtdt^{\prime}\nu_{A}(t)\bar{\nu}_{B}(t^{\prime})[a^{2}W_{\mathrm{BTZ}}^{M_{1}}(x_{A}(t),x_{B}(t^{\prime}))\notag
		\\&+b^{2}W_{\mathrm{BTZ}}^{M_{2}}(x_{A}(t),x_{B}(t^{\prime}))+2abW_{\mathrm{BTZ}}^{M_{1}M_{2}}(x_{A}(t),x_{B}(t^{\prime}))],\label{c}
		\end{align}
		\begin{align}
		\mathcal{M}=&-2\lambda^{2}\int dtdt^{\prime}\nu_{A}(t)\nu_{B}(t^{\prime})[(a^{2}+ab)W_{\mathrm{BTZ}}^{M_{1}}(x_{B}(t^{\prime}),x_{A}(t))\notag
		\\&+(b^{2}+ab)W_{\mathrm{BTZ}}^{M_{2}}(x_{B}(t^{\prime}),x_{A}(t))],\label{x}
		\end{align}
	where
		\begin{align}
		a=&\cos\theta\cos\varphi,\notag\\
           b=&\sin\theta\sin\varphi,\label{ab}
		\end{align}
		\begin{align}
		P_D^{M_i}=\int dtdt^{\prime}\nu_D(t)\bar{\nu}_D(t^{\prime})W_{\mathrm{BTZ}}^{M_i}(x_D(t),x_D(t^{\prime})),\label{pdmi}
		\end{align}
	and
		\begin{align}
		P_D^{M_1M_2}=\int dtdt^{\prime}\nu_D(t)\bar{\nu}_D(t^{\prime})W_{\mathrm{BTZ}}^{M_1M_2}(x_D(t^{\prime}),x_D(t))\label{pdm12}
		\end{align}
	with $\nu_D(t)=\frac{d\tau_\mathrm{D}}{dt}\Upsilon_D(\tau_D(t))e^{-\mathrm{i}\Omega_D\tau_D(t)}$ and $\Upsilon_D(\tau_D(t))=e^{-\tau_D(t)^2/2\sigma^2}$ is the Gaussian switching function.
	The off-diagonal elements $\mathcal{M}$ and $\mathcal{L}_{\mathrm{AB}}$ correspond to the nonlocal terms that depend on both trajectories, with $\mathcal{M}$ responsible for entangling the two detectors and $\mathcal{L}_{\mathrm{AB}}$ used for calculating the mutual information.
	Using (\ref{Wightman}) and (\ref{Wightmanspuer}), we can evaluate the matrix components, and for that we focus on two identical static detectors along the same axis at the black hole's center ($\Delta\phi=0$) and assume both detectors switch on and off simultaneously.
	With that, the matrix components up to order $\lambda^{2}$ are calculated numerically in Appendix \ref{appa2}.
	
\begin{figure}[h]
		\centering
		\includegraphics[scale=0.41]{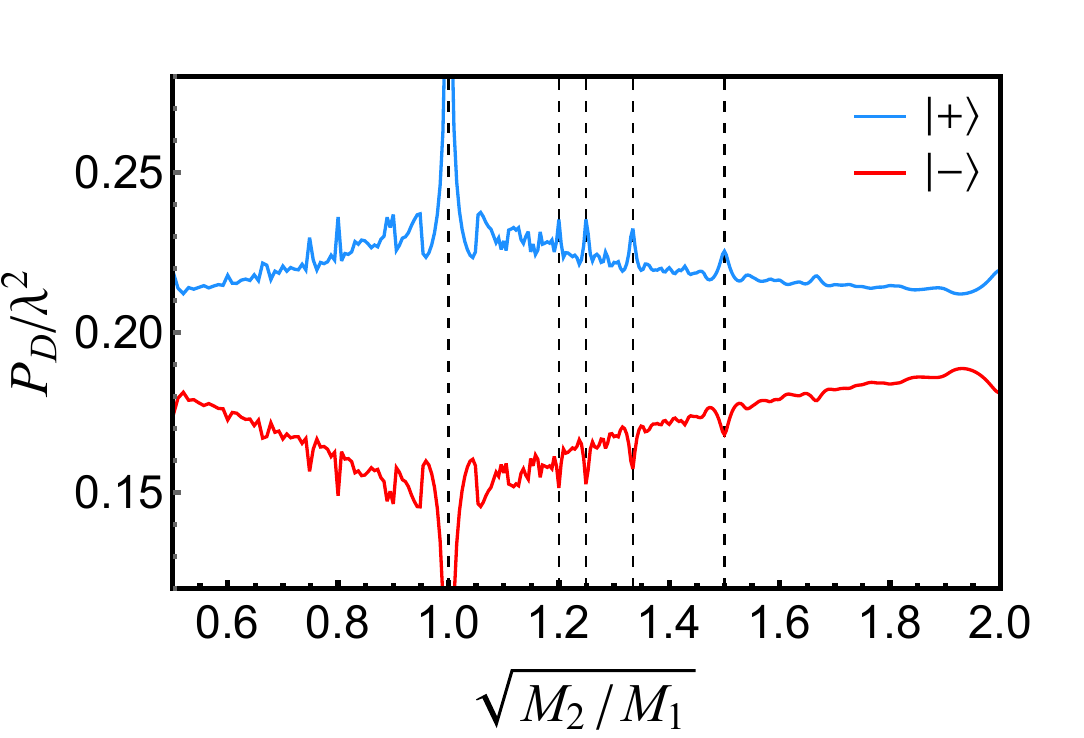}
		\caption{ Transition probability of the individual detector as a function of $\sqrt{M_2/M_1}$ with parameters $l=10\sigma$, $r_D=10\sigma$, $\sigma\Omega=0.01$, and $\zeta=1$.
		The measurement basis corresponding to the relevant plot is indicated by the legend.
The dashed vertical lines are at rational values of $\sqrt{M_2/M_1}$ (e.g. 1, 1.2, 1.5).}
		\label{fig1}
		\end{figure}

	It is important to note that the state (\ref{matrix}) is not normalized, since we are considering the final conditional state of the detector.
	To have a normalized density matrix, one must divide it with $\mathrm{Tr}(\rho_{AB})$ (see Appendix \ref{appa3}), and by doing so we land up with the following density operator
		\begin{equation}
		\rho_{AB}=
		\begin{pmatrix}
		\tilde{P}_{G}&0&0&\mathcal{M}\\0&P_B&\mathcal{L}_{\mathrm{AB}}&0\\0&\mathcal{L}_{\mathrm{AB}}^*&P_A&0\\\mathcal{M}^*&0&0&0
		\end{pmatrix}+\mathcal{O}(\lambda^{4}),\label{matrixnor}
		\end{equation}
	where $\tilde{P}_{G}=1-P_A-P_B$.
	Thus we can find the reduced state of detector $A$ or $B$ as
		\begin{equation}
		\rho_A=\mathrm{Tr}_B(\rho_{AB})=
		\begin{pmatrix}1-P_A&0\\0&P_A
		\end{pmatrix}
		\end{equation}
	and vice versa.
	At this point, one can verify this density matrix corresponds to that derived in \cite{Foo:2021exb}, which examined a single detector in a superposed BTZ spacetime background.
	In particular, we discuss two special cases as shown in Fig. \ref{fig1}:
	
	(1) Setting $\theta=\varphi=\frac{\pi}{4}$ prepares the initial spacetime state and final measurement control state as $|s_i\rangle=|s_{f}\rangle \rightarrow|+\rangle=\frac{1}{\sqrt{2}}(|M_1\rangle+|M_2\rangle)$, a symmetric superposition of $|M_1\rangle$ and $|M_2\rangle$, yielding the transition
probability $P_{D}^{+}/\lambda^2=\frac{1}{4}(P_{D}^{M_{1}}+P_{D}^{M_{2}}+2P_{D}^{M_{1}M_{2}})$ as calculated in \cite{Foo:2021exb}.

	(2) Setting $\theta=\frac{\pi}{4}$ and $\varphi=-\frac{\pi}{4}$ prepares the initial spacetime state in $|+\rangle$, while the final spacetime state is an antisymmetric superposition, $|s_f\rangle\to|-\rangle=\frac{1}{\sqrt{2}}(|M_1\rangle-|M_2\rangle)$.
	This yields the transition probability $P_{D}^{-}/\lambda^2=\frac{1}{4}(P_{D}^{M_{1}}+P_{D}^{M_{2}}-2P_{D}^{M_{1}M_{2}})$ as calculated in \cite{Foo:2021exb}.
		\noindent	
		
	In Fig. \ref{fig1}, the transition probability of the single detector is plotted as a function of the values of the square root ratio of superposed masses$\sqrt{M_2/M_1}$.
	As predicted, resonant peaks arise at rational values of the square root ratio of superposed masses, driven by constructive interference among field modes in topologically closed AdS spacetimes, yielding resonances in the detector response at integer values of $\sqrt{M_2/M_1}$.
	As $\sqrt{M_2/M_1}\rightarrow1$, the transition probability approaches that of a single black hole for a measurement in $|+\rangle$, whereas it vanishes for $|-\rangle$.
	This is not difficult to understand, because the two masses are superposed at the same center point, as explained explicitly in \cite{Foo:2021exb}. 
	When two equal masses are superposed in a specific proportion, the resultant black hole is equivalent to a classical black hole of the same mass.
	
\section{RESULTS} \label{sec4}
	To quantify the effect of superposed spacetime on entanglement extraction following the interaction of spacelike-separated detectors with the field, we employ concurrence as the measure of entanglement, which, with the density matrix (\ref{matrixnor}) is \cite{Hill:1997pfa,Wootters:1997id}
		\begin{equation}
		\mathcal{C}(\rho_{AB})=2\mathrm{Max}\left[0,\left(|\mathcal{M}|-\sqrt{P_{A}P_{B}}\right)\right]+O(\lambda^{4}).\label{concu}
		\end{equation}
	Clearly, the concurrence is a competition between the correlation term $\mathcal{M}$ and the detector's transition probabilities $P_{A}$ and $P_{B}$.
	
	The total correlations, encompassing both classical and quantum contributions, between detectors are quantified by mutual information, defined as follows
		\begin{align}
		\mathcal{I}(\rho_{AB})=&\mathcal{L}_{+}\ln\mathcal{L}_{+}+\mathcal{L}_{-}\ln\mathcal{L}_{-}\notag\\
		&-P_A\ln P_A-P_B\ln P_B+O(\lambda^{4}),\label{mutaul}
		\end{align}
	with
		\begin{equation}
		\mathcal{L}_{\pm}:=\frac{1}{2}\left[P_A+P_B\pm\sqrt{\left(P_A-P_B\right)^{2}+4\left|\mathcal{L}_{\mathrm{AB}}\right|^{2}}\right].\label{mutaul2}
		\end{equation}
	Note that the mutual information $\mathcal{I}(\rho_{AB})$ is determined by the transition probabilities $P_A$ and $P_B$, and the correlation term $\mathcal{L}_{\mathrm{AB}}$.
	Furthermore, mutual information remains nonzero even when the concurrence $C(\rho_{AB})=0$, indicating that the extracted correlations between detectors are either classical or non-distillable entanglement.
	
\subsection{Entanglement harvesting}\label{sub1}
	When a pair of UDW detectors individually couple to the quantum field, their initial state, together with the field, forms a product state.
	Following their interaction, correlations between the detectors emerge, attributable either to direct field-mediated communication or to correlations swapping from the vacuum, particularly when the detectors are spacelike separated, precluding direct communication due to causality constraints.
	\begin{figure}[h]
		\centering
		\includegraphics[scale=0.35]{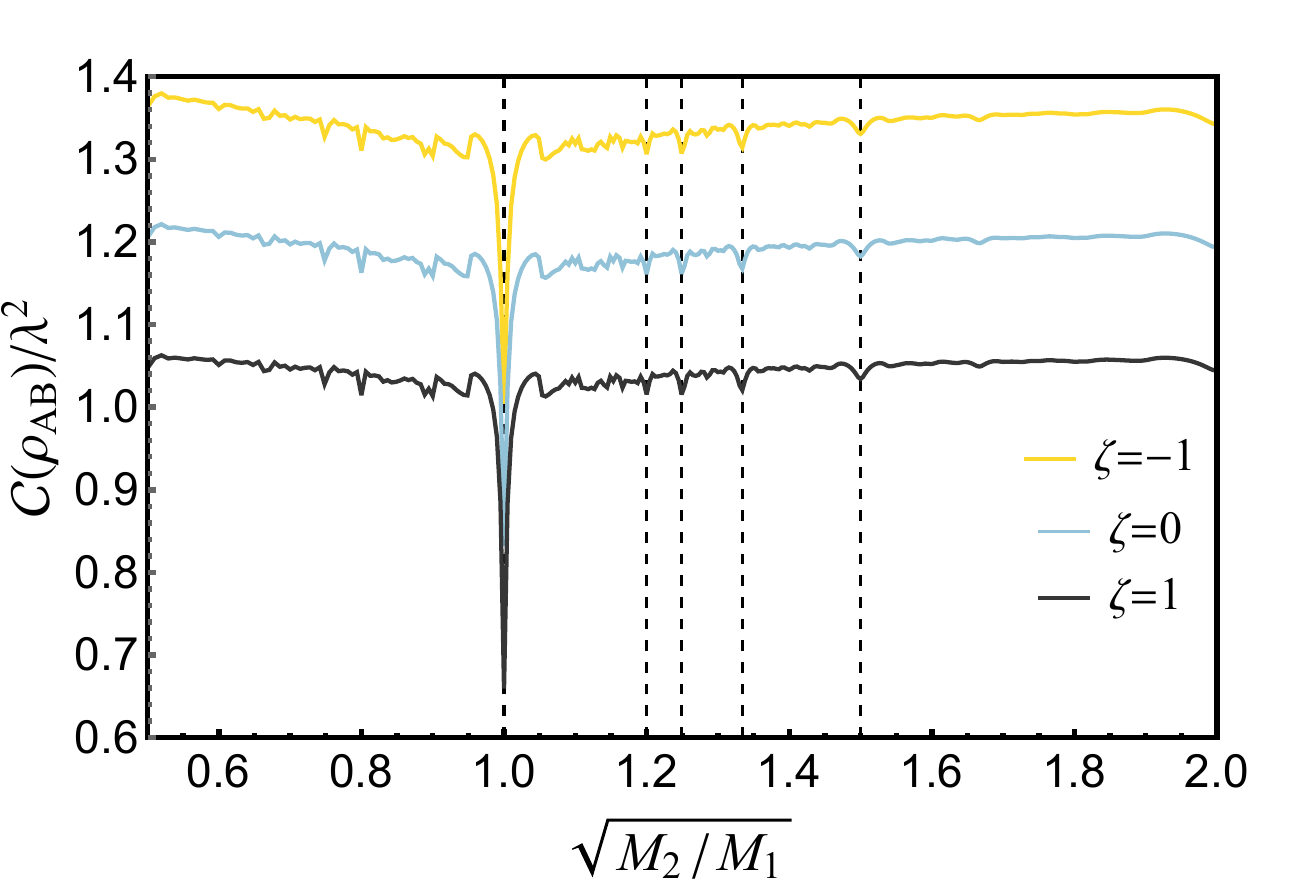}
		\caption{ The concurrence $\mathcal{C}(\rho_{AB})/\lambda^{2}$ between two detectors in the superposed BTZ spacetime as a function of $\sqrt{M_2/M_1}$ is plotted for different values of $\zeta$ with parameters $l=10\sigma$, $r_A=10\sigma$, $r_B=11\sigma$, $\sigma\Omega=0.01$, and $\theta=\varphi=\frac{\pi}{4}$. Vertical dashed lines were added for distinctive peaks at rational values (e.g. 1, 1.2, 1.5).}
		\label{fig2}
		\end{figure}\noindent	

\begin{widetext}
	
	\begin{figure}[h]
		\centering
		\includegraphics[scale=0.34]{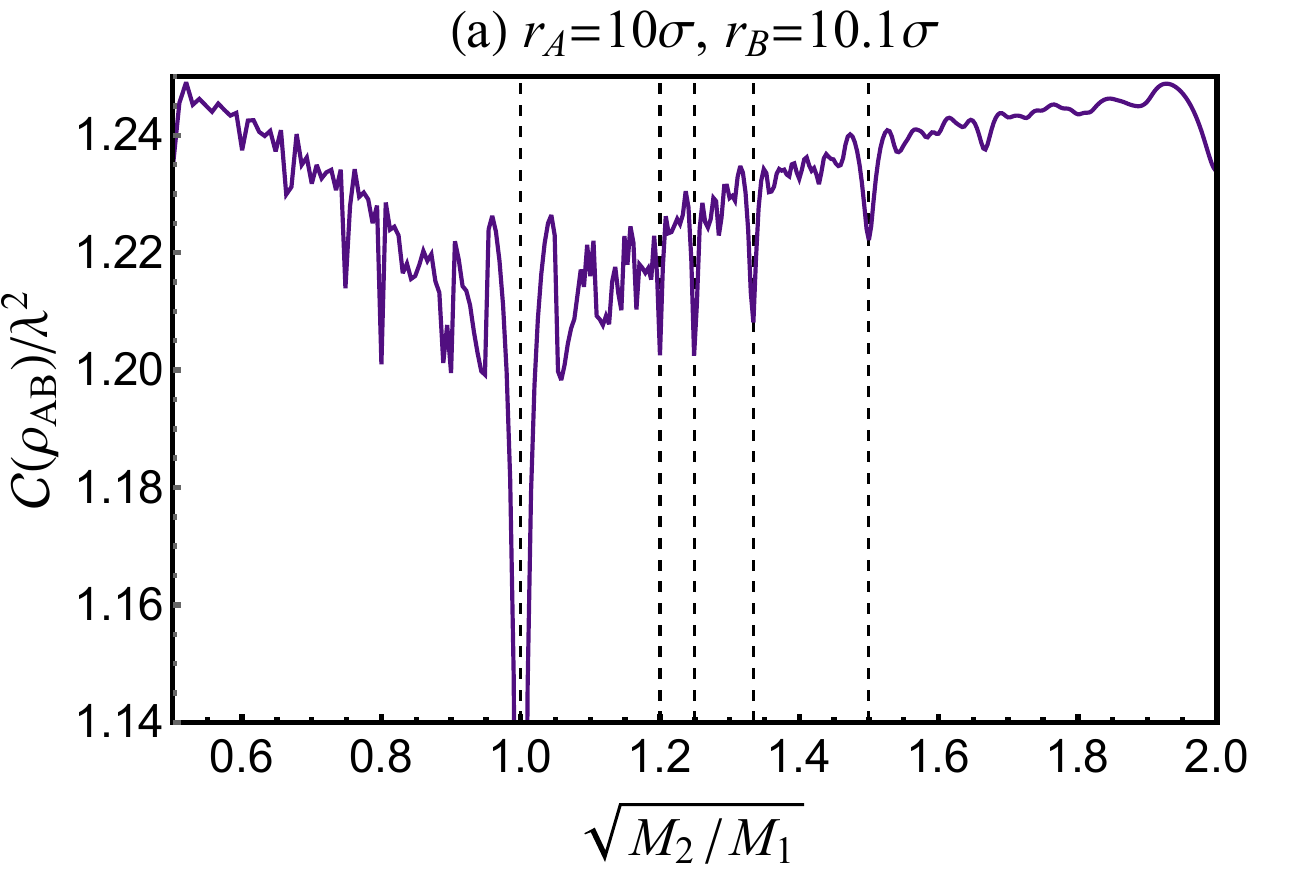}
		\includegraphics[scale=0.34]{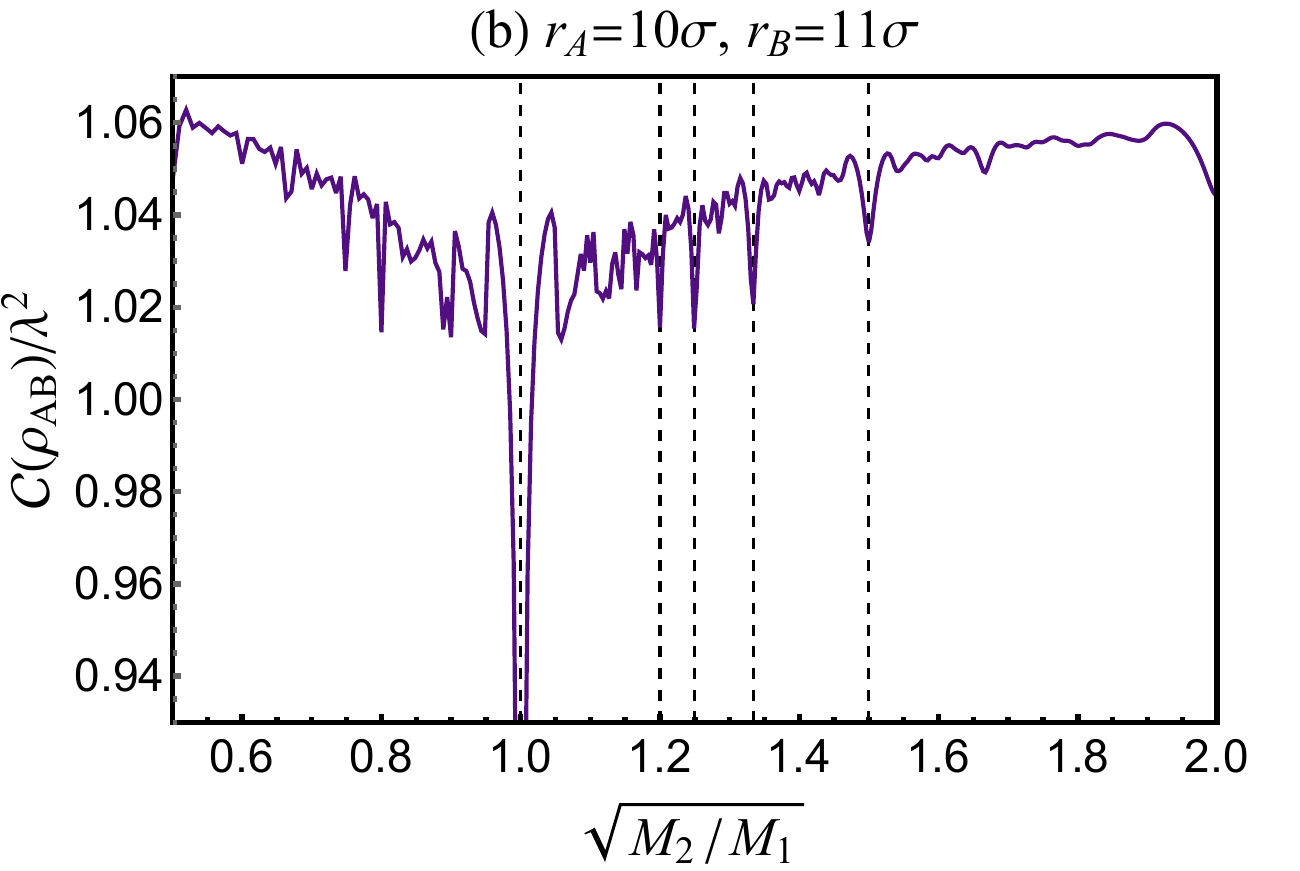}
		\includegraphics[scale=0.34]{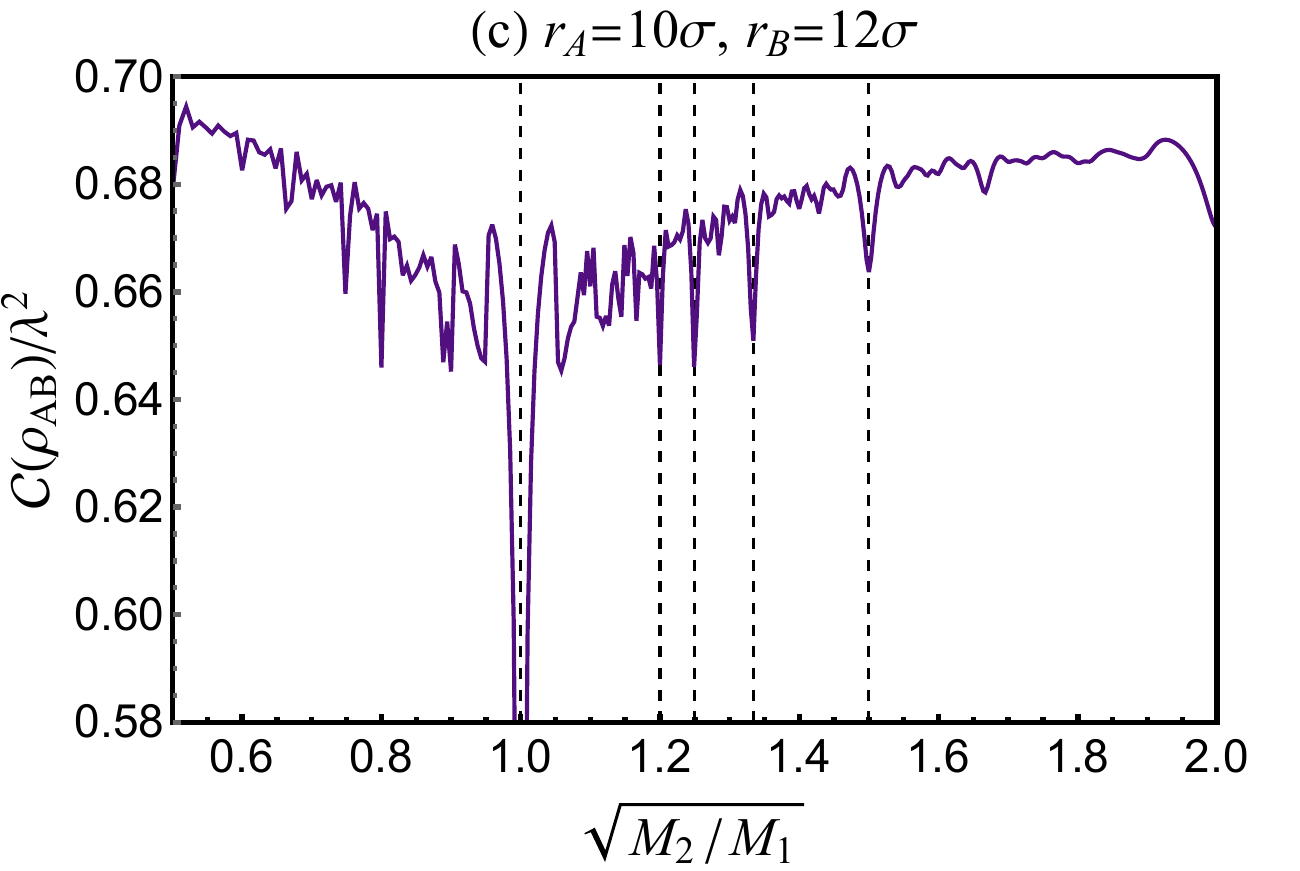}
		\includegraphics[scale=0.34]{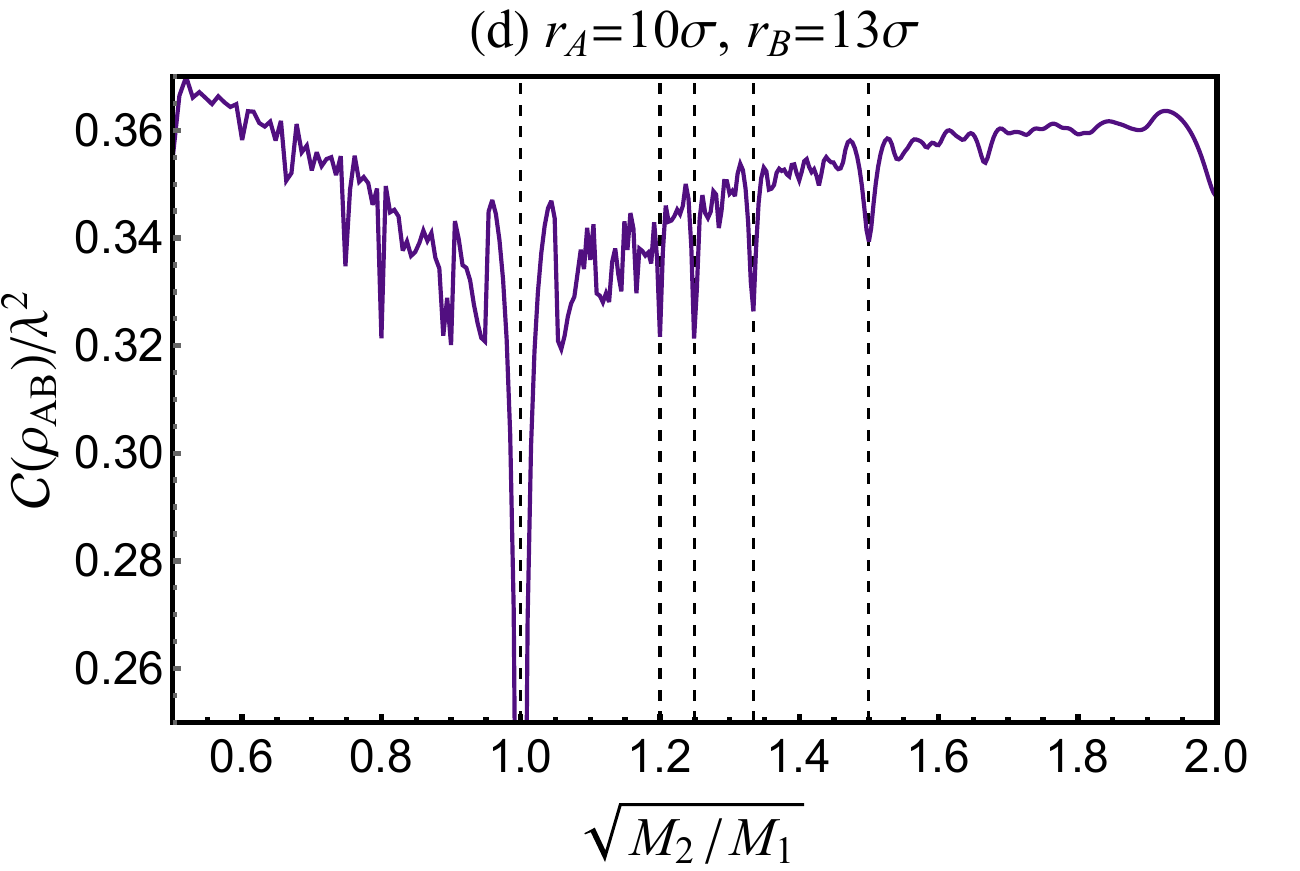}
		\caption{ The concurrence $\mathcal{C}(\rho_{AB})/\lambda^{2}$ between two detectors in the superposed BTZ spacetime as a function of $\sqrt{M_2/M_1}$ is plotted for different values of $r_B$ with parameters $l=10\sigma$, $\sigma\Omega=0.01$, and $\theta=\varphi=\frac{\pi}{4}$.}
		\label{fig3}
		\end{figure}\noindent	
	\end{widetext}

	We now consider harvested entanglement in the case of the detectors subjected to the superposition of spacetime when the detectors A and B are located at fixed values $r_A$ and $r_B$ of the BTZ radial coordinate $r$.
	The transition probabilities $P_A$ and $P_B$, as well as the matrix element $\mathcal{M}$, may be obtained numerically using (\ref{pd}) and (\ref{x}), after which the concurrence given by (\ref{concu}) can be easily assessed for the generation of entanglement between the detectors.

	In Fig. \ref{fig2}, the amount of obtained entanglement is plotted as a function of $\sqrt{M_2/M_1}$.
	Here, we show that quantum signatures can also be observed through the entanglement harvesting process, which corroborates the results of \cite{Foo:2021exb}.
	Furthermore, the amplitude of the oscillation decreases in the limit of $\sqrt{M_2/M_1}\ll1$ or $\sqrt{M_2/M_1}\gg1$ due to the decay of the correlation term $P_{D}^{M_{1}M_{2}}$.
	\begin{figure}[h]
		\centering
		\includegraphics[scale=0.34]{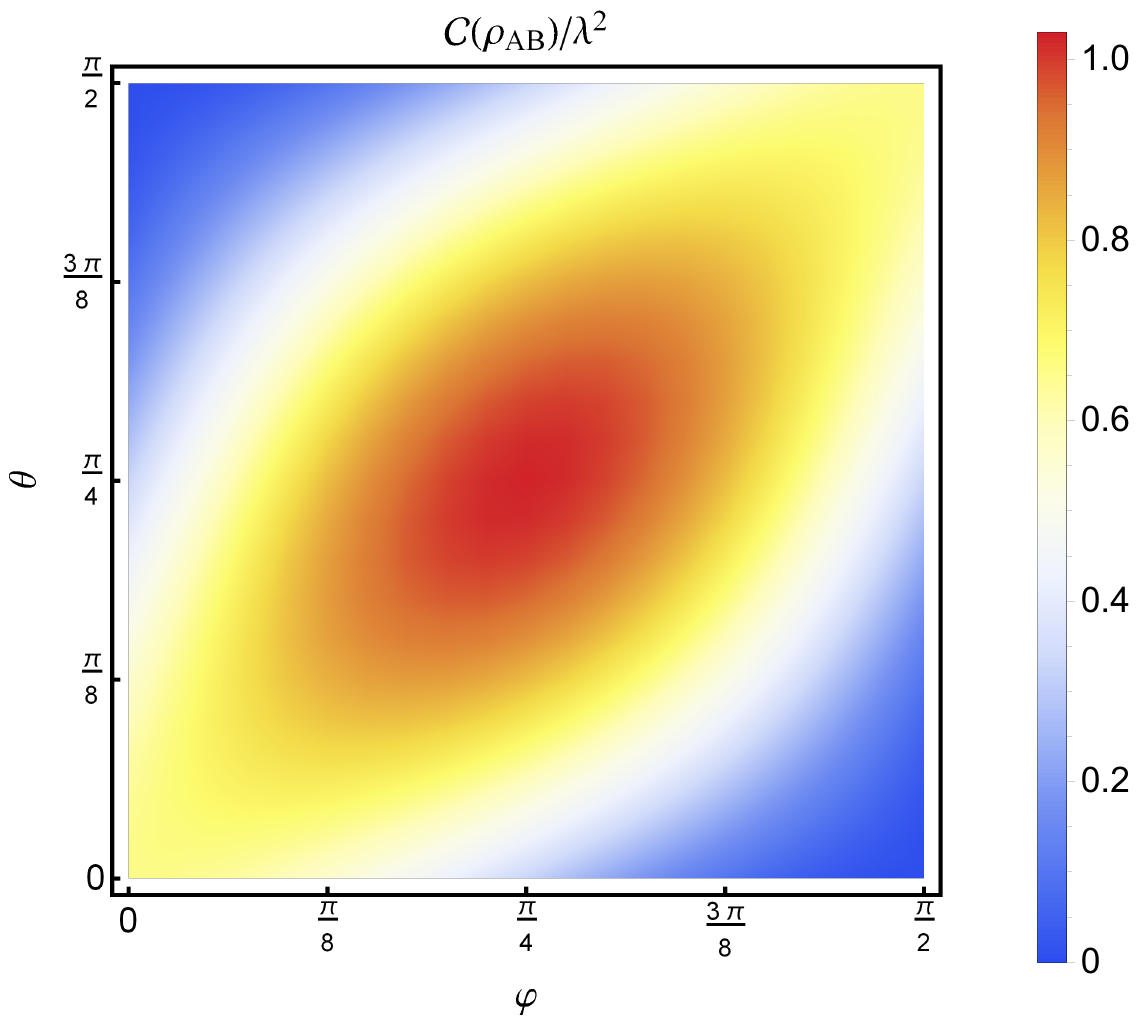}
		\caption{The concurrence $\mathcal{C}(\rho_{AB})/\lambda^2$ for superposed spacetime, plotted as a function of $\theta$ and $\varphi$, exhibits a maximum near $\theta = \varphi$.
		We use parameters $l=10\sigma$, $\sqrt{M_2/M_1} =1.5$, $r_A=10\sigma$, $r_B=11\sigma$, and $\sigma\Omega=0.01$.}
		\label{fig4}
		\end{figure}\noindent	
	It is noteworthy that the amount of entanglement harvested is always greater in the superposed spacetime compared to single spacetime.
	This behavior is intuitively anticipated; the quantum field mode constructive interference in superposed spacetimes markedly enhances entanglement harvesting relative to a single spacetime background.
	In order to facilitate comparison with \cite{Henderson:2017yuv,Bueley:2022ple,Liu:2025bpp}, we will choose the Dirichlet boundary condition $\zeta=1$ throughout the following article.
	
	Next, to see the effects purely coming from the black hole, we consider the influence of different proper distances between the two detectors on entanglement harvesting.
	As shown in Fig. \ref{fig3}, as the separation between the detectors grows, the entanglement between the detectors decreases.
	This behavior is as anticipated since correlations in the vacuum state diminish significantly for spacetime points separated by a large distance, which can be seen from the BTZ Wightman function in (\ref{Wightmanspuer}).	
Moreover, the findings presented in Fig. \ref{fig3} are consistent with the results reported in \cite{Henderson:2017yuv,Liu:2025bpp}.
	
	Finally, in order to further investigate the impact of spacetime superposition on entanglement harvesting, we analyze the concurrence behavior, as depicted in Fig.  \ref{fig4}, with respect to variations in $\theta$ and $\varphi$, which parameterize the initial and final quantum states of the spacetime.
	We show that maximal entanglement is attained when $\theta = \varphi$, corresponding to the condition where the control state of the spacetime is measured to be identical to the initially prepared spacetime state.

\subsection{Mutual information harvesting}\label{sub2}
	Mutual information \cite{Mutual} quantifies the total of classical and quantum correlations, including entanglement.

	\begin{widetext}
	
	\begin{figure}[h]
		\centering
		\includegraphics[scale=0.35]{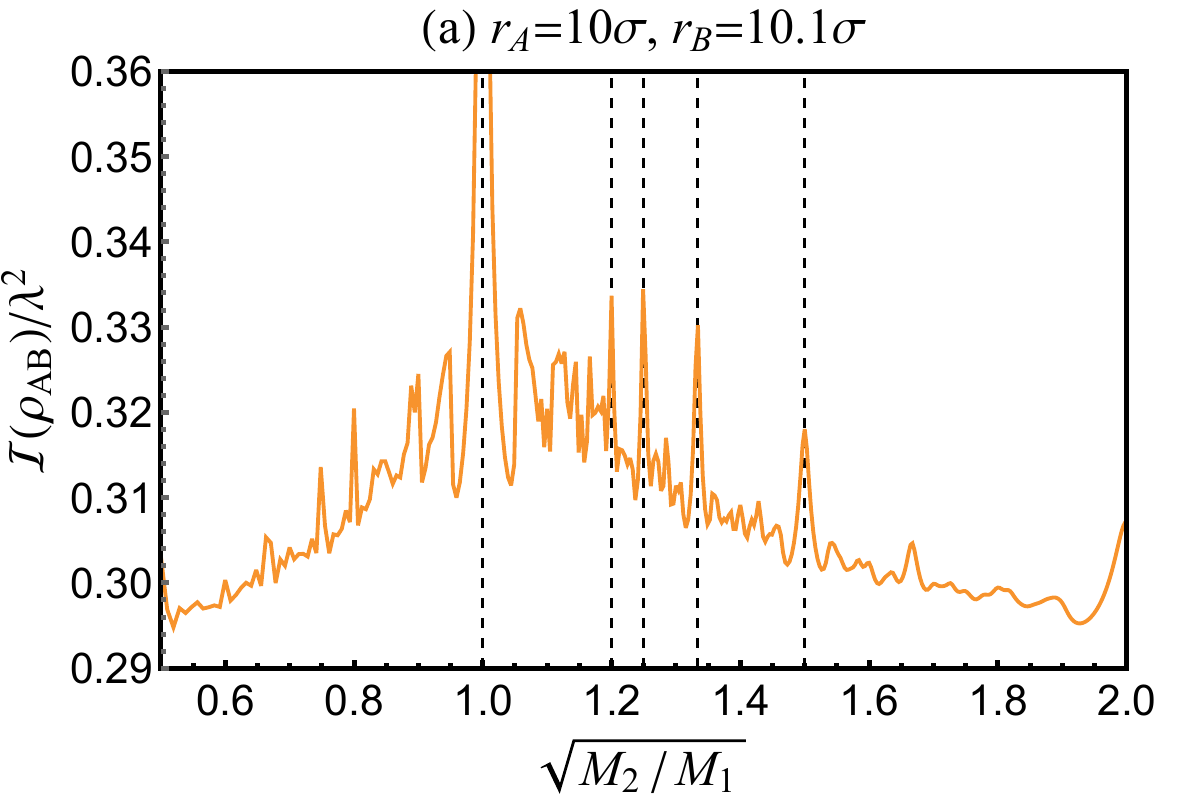}
		\includegraphics[scale=0.35]{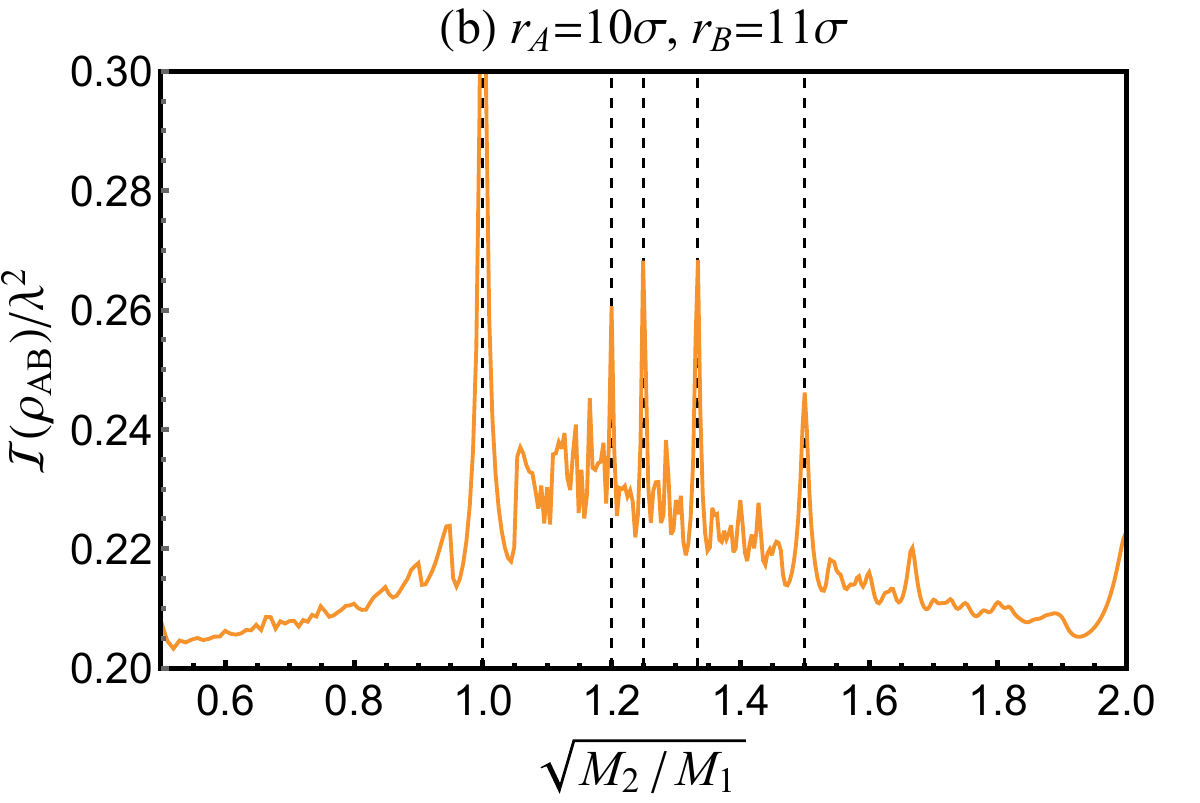}
		\includegraphics[scale=0.35]{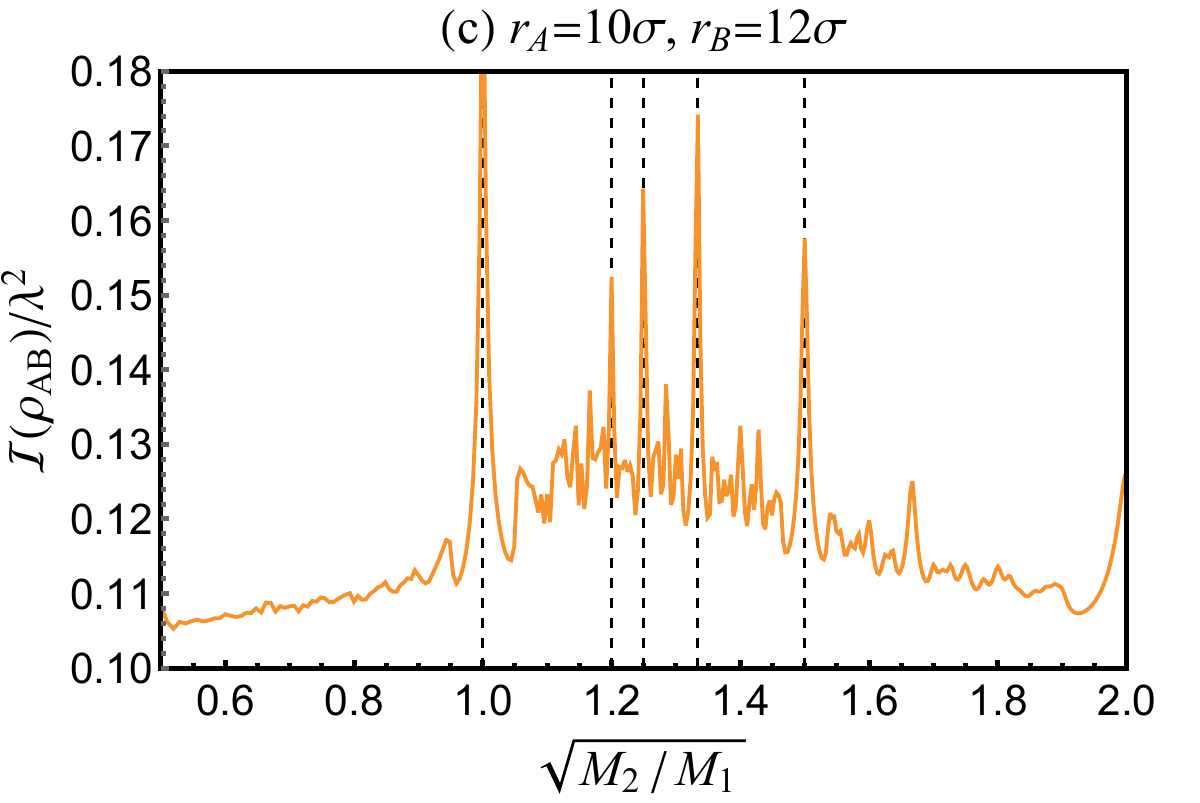}
		\includegraphics[scale=0.35]{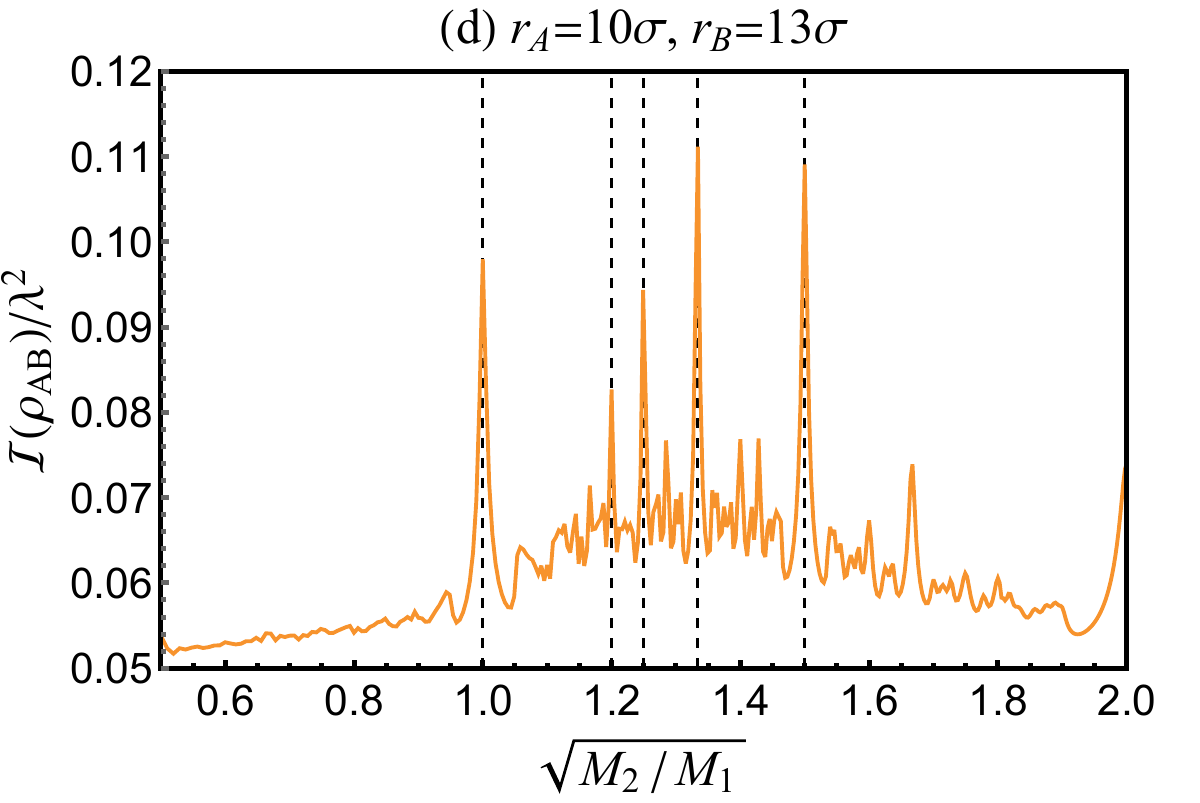}
		\caption{ The mutual information $\mathcal{I}(\rho_{AB})/\lambda^{2}$ between two detectors in the superposed BTZ spacetime as a function of $\sqrt{M_2/M_1}$ is plotted for different values of $r_B$ with parameter $l=10\sigma$, $\sigma\Omega=0.01$, and $\theta=\varphi=\frac{\pi}{4}$.}
		\label{fig5}
		\end{figure}\noindent	
	\end{widetext}
	
	By examining mutual information and entanglement concurrently, we can investigate critical distinctions between the impacts of superposition of spacetime on classical and quantum correlations.
	Now, we begin to numerically evaluate the mutual information harvesting between detectors as given in (\ref{mutaul}).
	\begin{figure}[h]
		\centering
		\includegraphics[scale=0.34]{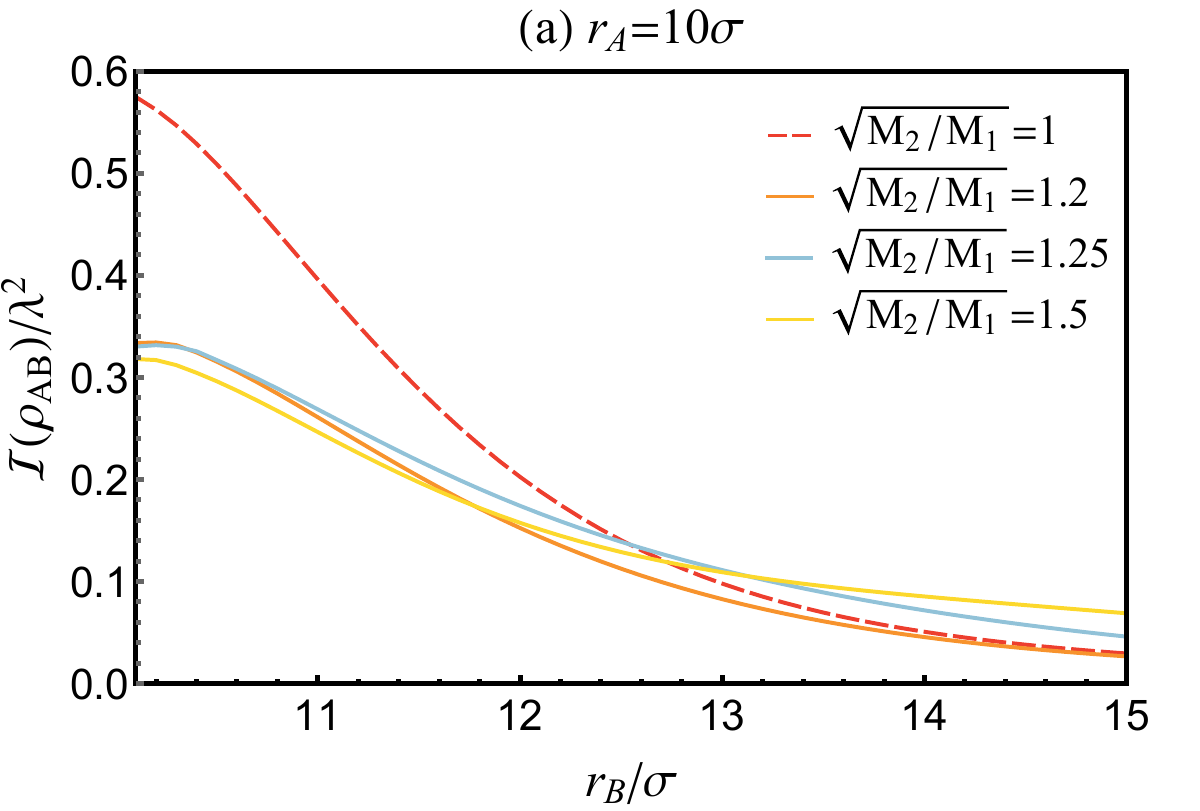}
		\includegraphics[scale=0.34]{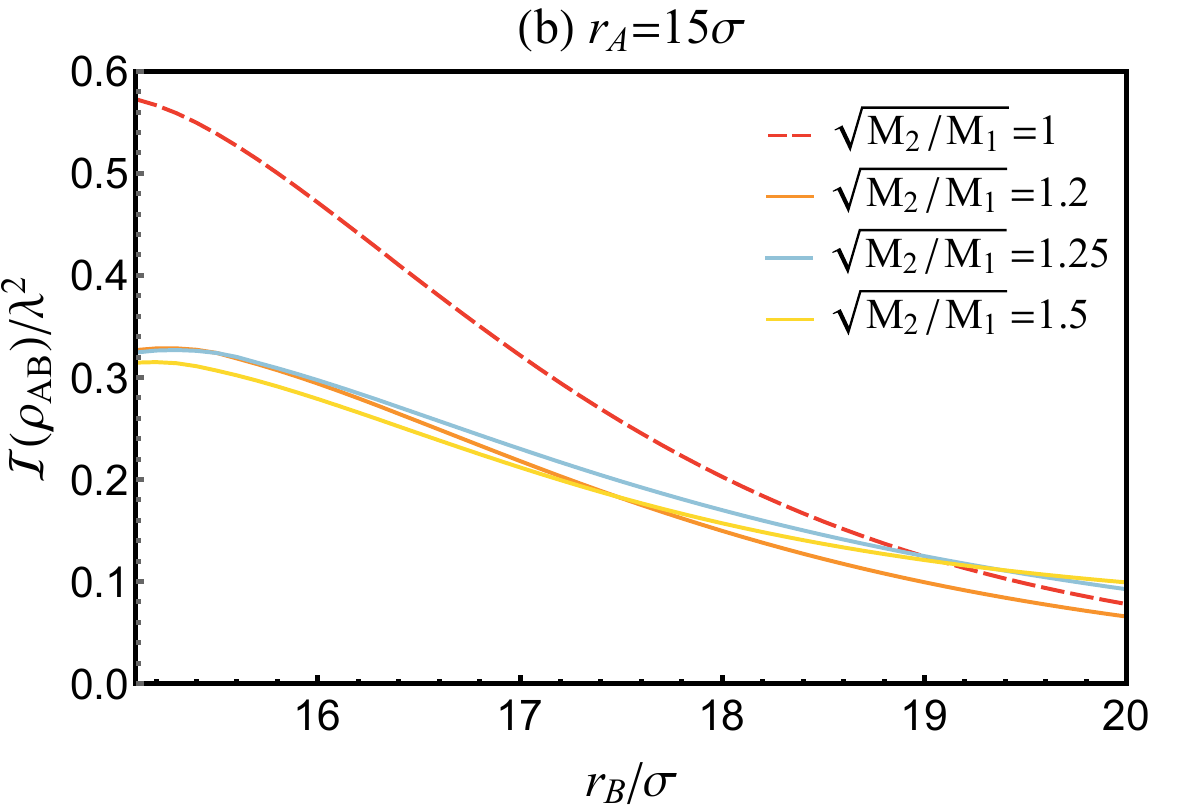}
		\caption{The mutual information $\mathcal{I}(\rho_{AB})/\lambda^{2}$ for superposed spacetime as a function of $r_B$ is plotted for different values of $\sqrt{M_2/M_1}$ with parameters $l=10\sigma$, $\sigma\Omega=0.01$, and $\theta=\varphi=\frac{\pi}{4}$.}
		\label{fig6}
		\end{figure}\noindent	
		\begin{figure}[h]
		\centering
		\includegraphics[scale=0.36]{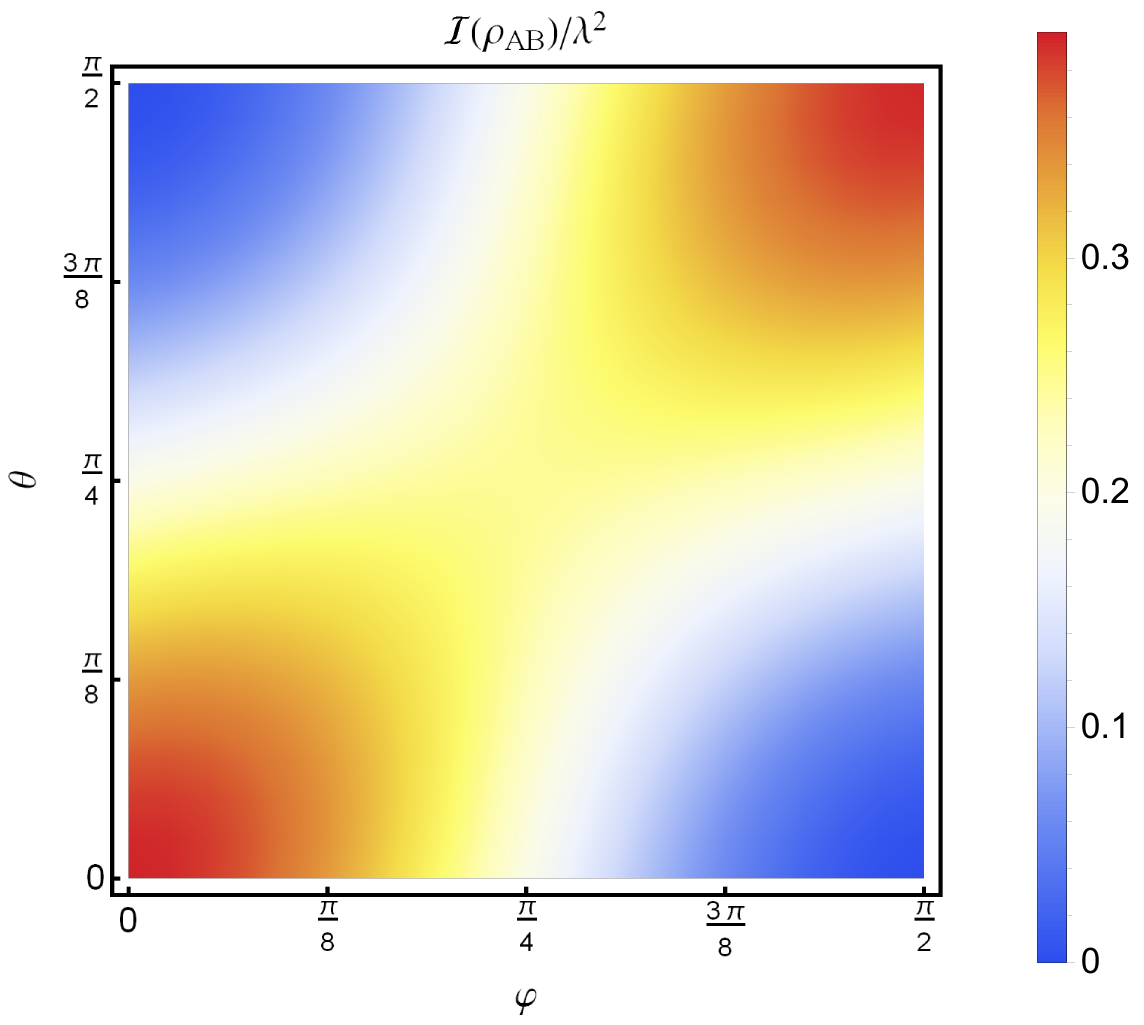}
		\caption{The mutual information $\mathcal{I}(\rho_{AB})/\lambda^{2}$ for superposed spacetime, plotted as a function of $\theta$ and $\varphi$, exhibits a maximum near $\theta = \varphi$.
		We use parameters $l=10\sigma$, $\sqrt{M_2/M_1} =1.5$, $r_A=10\sigma$, $r_B=11\sigma$, and $\sigma\Omega=0.01$.}
		\label{fig7}
		\end{figure}\noindent	
	Fig. \ref{fig5} show the amount of mutual information harvesting as a function of $\sqrt{M_2/M_1}$ for various proper distance between the two detectors.
	First, similar to entanglement, the mutual information between detectors diminishes as their separation increases.
	However, mutual information harvested in superposed spacetime is more sensitive to the distance between detectors.
	When the proper distance between detectors is small, the mutual information harvested in a superposed spacetime is consistently less than that in a single spacetime.
	As the separation of the detectors increases, the mutual information harvested in a superposed spacetime surpasses that in a single spacetime for specific mass ratios $\sqrt{M_2/M_1}$.
 This differs from entanglement harvested in superposed spacetime, where the amount of entanglement harvested always exceeds that in a single spacetime, regardless of the distance between the detectors.
	To investigate the correlation between mutual information harvesting and the proper distance of detectors in the background of superposed spacetime, we present Fig.   \ref{fig6} and find that with increasing detector separation, the mutual information extracted in a superposed spacetime surpasses that in a single spacetime for specific mass ratios $\sqrt{M_2/M_1}$.
	Similarly, we also examine the nature of mutual information with varying $\theta$ and $\varphi$ in Fig. \ref{fig7}.
	Analogous to entanglement, maximal mutual information is achieved in the region where $\theta = \varphi$, which implies that the detector achieves maximal mutual information when the measured spacetime control state precisely matches the initially prepared spacetime state.

\section{CONCLUSIONS AND OUTLOOKS} \label{sec5}
	We investigate the correlation harvesting for a spacetime in quantum superposition.
	In particular, we employ an operational framework to analyze `superpositions of spacetimes' by coupling matter, represented by an UDW detector, to a quantum field.
	We utilize this methodology to investigate the entanglement and mutual information harvested in a BTZ black hole in superposition.
	We have shown that the correlation harvesting process reveals signatures indicative of quantum superposition.
	This effect confirms a related result obtained recently for the (2+1)-dimensional BTZ black hole in a superposition of masses \cite{Foo:2021exb}.

	In the framework of entanglement harvesting, the constructive interference between the field modes in superposed spacetime significantly enhances the extractable quantum entanglement.
	This study demonstrates that, while entanglement harvesting is predominantly influenced by the local properties of the quantum field vacuum, the global spacetime structure, particularly its superposed configuration, significantly enhances quantum entanglement through induced interference effects.
	In contrast to entanglement harvesting, the effect of spacetime superposition on mutual information harvesting is influenced by the proper distance between the two detectors.
	For small detector separations, the mutual information harvested in a superposed spacetime is consistently lower than in a single spacetime; however, as the separation increases, it surpasses that of a single spacetime for specific mass ratios $\sqrt{M_2/M_1}$.
	Our study further finds that maximal entanglement and mutual information are achieved when the measured control state of the spacetime coincides with its initially prepared state.
	
	Our results elucidate the impact of quantum gravitational phenomena, specifically spacetime superposition, on relativistic quantum information processing, such as entanglement and mutual information harvesting.
	Given that quantum discord \cite{Ollivier:2001fdq,Henderson:2001wrr} quantifies nonclassical correlations, future studies may explore the effects of spacetime superposition on quantum discord harvesting, elucidating the distinct influences of spacetime superposition on entanglement harvesting and mutual information harvesting.
	On the other hand, while our study focused on the relatively simple framework of mass-superposed BTZ spacetime, future research may investigate superpositions of more complex curved spacetimes or dynamic scenarios involving temporally evolving superposed geometries.
	Such investigations could elucidate the behavior of quantum fields and relativistic quantum information in more realistic spacetime models, thereby advancing our understanding of quantum gravity.

\acknowledgments
This work was supported by the National Natural Science Foundation of China under Grants No.12475051, No.12374408, and No.12035005; the science and technology innovation Program of Hunan Province under grant No. 2024RC1050; the innovative research group of Hunan Province under Grant No. 2024JJ1006; and the Natural Science Foundation of Hunan Province under grant No. 2023JJ30384.

\begin{widetext}	
\appendix
\section{CALCULATION OF THE COMPONENTS OF REDUCED DENSITY MATRIX} \label{appa1}
	
	The unitary operators $U^{(n)}$s up to order of $\lambda^{2}$ in (\ref{U2}) are given by
		\begin{align}
		U^{(0)}=&\mathbf{I}\notag\\
		U^{(1)}=&-i\int dt\left[\left(\frac{d\tau_A}{dt}\right)H_A^I(\tau_A(t))\otimes\mathbf{I}+\mathbf{I}\otimes\left(\frac{d\tau_B}{dt}\right)H_B^I(\tau_B(t))\right]\notag\\
		U^{(2)}=&-\hat{\mathcal{T}}\int dt\int dt'\Bigg[\frac{d\tau_{A}}{dt}\frac{d\tau_{A}}{dt'}H_{A}^I(\tau_{A}(t))H_{A}^I(\tau_{A}(t'))\otimes\mathbf{I}+\mathbf{I}\otimes\frac{d\tau_{B}}{dt}\frac{d\tau_{B}}{dt'}H_{B}^I(\tau_{B}(t))H_{B}^I(\tau_{B}(t'))\notag\\
		&+\frac{d\tau_A}{dt}H_A^I(\tau_A(t))\otimes\frac{d\tau_B}{dt^{\prime}}H_B^I(\tau_B(t^{\prime}))+\frac{d\tau_A}{dt^{\prime}}H_A^I(\tau_A(t^{\prime}))\otimes\frac{d\tau_B}{dt}H_B^I(\tau_B(t))\Bigg].\label{U22}
		\end{align}
	The associated states $|\Phi_f^{(n)}\rangle$s, up to second order in $\lambda$, are expressed as follows
		\begin{align}
		|\Phi_{f}^{(0)}\rangle=&|\Phi_{i}\rangle\notag\\
		|\Phi_{f}^{(1)}\rangle=&-i\int_{-\infty}^{\infty}dt[|1,0\rangle\otimes\bar{\nu}_A(t)\{\cos\theta\hat{\phi}^{M_{1}}(x_{A})|0\rangle_{F}\otimes|M_{1}\rangle+\sin\theta\hat{\phi}^{M_{2}}(x_{A})|0\rangle_{F}\otimes|M_{2}\rangle\}+|0,1\rangle\otimes(A\leftrightarrow B)],\notag\\
		|\Phi_{f}^{(2)}\rangle=&-\int_{-\infty}^{\infty}dt\int_{-\infty}^{t}dt^{\prime}[|0,0\rangle\otimes\{\nu_{A}(t)\bar{\nu}_{A}(t^{\prime})(\cos\theta\hat{\phi}^{M_{1}}(x_{A})\hat{\phi}^{M_{1}}(x_{A}^{\prime})|0\rangle_{F}\otimes|M_{1}\rangle\notag\\
		&+\sin\theta\hat{\phi}^{M_{2}}(x_{A})\hat{\phi}^{M_{2}}(x_{A}^{\prime})|0\rangle_{F}\otimes|M_{2}\rangle)+(A\leftrightarrow B)\}+2|1,1\rangle\otimes\bar{\nu}_{A}(t)\bar{\nu}_{B}(t^{\prime})\notag\\
		&\times(\cos\theta\hat{\phi}^{M_{1}}(x_{A})\hat{\phi}^{M_{1}}(x_{B}^{\prime})|0\rangle_{F}\otimes|M_{1}\rangle+\sin\theta\hat{\phi}^{M_{2}}(x_{A})\hat{\phi}^{M_{2}}(x_{B}^{\prime})|0\rangle_{F}\otimes|M_{2}\rangle)],\label{phi}
		\end{align}
	where $|0,0\rangle\equiv|0\rangle_A\otimes|0\rangle_B$, $\nu_A(t)=\frac{d\tau_{A}}{dt}\Upsilon_{A}(\tau_{A})e^{-\mathrm{i}\Omega_D\tau_D(t)}$ and $\hat{\phi}^{M_{i}}(x_{D})\equiv\hat{\phi}^{M_{i}}(x_{D}(\tau_{D}(t)))$.
	By tracing over all possible field configurations and performing measurements in the final spacetime control state $\left|s_f\right\rangle$, the components of the joint detector state can be determined, which is given by
		\begin{equation}
		\mathrm{Tr}_\phi[\langle s_f|\Phi_f^{(0)}\rangle\langle\Phi_f^{(0)}|s_f\rangle]=(a+b)^2|0,0\rangle\langle0,0|,\label{fphi00}
		\end{equation}
		\begin{align}
		\mathrm{Tr}_{\phi}[\langle& s_{f}|\Phi_{f}^{(1)}\rangle\langle\Phi_{f}^{(1)}|s_{f}\rangle]\notag\\
		=&|1,0\rangle\langle1,0|\int dtdt^{\prime}\bar{\nu}_{A}(t)\nu_{A}(t^{\prime})[a^{2}W_{\mathrm{BTZ}}^{M_{1}}(x_{A}^{\prime},x_{A})+b^{2}W_{\mathrm{BTZ}}^{M_{2}}(x_{A}^{\prime},x_{A})+2abW_{\mathrm{BTZ}}^{M_{1}M_{2}}(x_{A}^{\prime},x_{A})]\notag\\
		+&|0,1\rangle\langle0,1|\int dtdt^{\prime}\bar{\nu}_{B}(t)\nu_{B}(t^{\prime})[a^{2}W_{\mathrm{BTZ}}^{M_{1}}(x_{B}^{\prime},x_{B})+b^{2}W_{\mathrm{BTZ}}^{M_{2}}(x_{B}^{\prime},x_{B})+2abW_{\mathrm{BTZ}}^{M_{1}M_{2}}(x_{B}^{\prime},x_{B})]\notag\\
		+&|1,0\rangle\langle0,1|\int dtdt^{\prime}\bar{\nu}_{A}(t)\nu_{B}(t^{\prime})[a^2W_{\mathrm{BTZ}}^{M_1}(x_B^{\prime},x_A)+b^2W_{\mathrm{BTZ}}^{M_2}(x_B^{\prime},x_A)+2abW_{\mathrm{BTZ}}^{M_1M_2}(x_B^{\prime},x_A)]\notag\\
		+&|0,1\rangle\langle1,0|\int dtdt^{\prime}\bar{\nu}_{B}(t)\nu_{A}(t^{\prime})[a^{2}W_{\mathrm{BTZ}}^{M_{1}}(x_{A}^{\prime},x_{B})+b^{2}W_{\mathrm{BTZ}}^{M_{2}}(x_{A}^{\prime},x_{B})+2abW_{\mathrm{BTZ}}^{M_{1}M_{2}}(x_{A}^{\prime},x_{B})],\label{fphi11}
		\end{align}
		\begin{align}
		\mathrm{Tr}_{\phi}[\langle& s_{f}|\Phi_{f}^{(2)}\rangle\langle\Phi_{f}^{(0)}|s_{f}\rangle]\notag\\
		=&-|0,0\rangle\langle0,0|(a+b)\left[\int dtdt^{\prime}\nu_{A}(t)\bar{\nu}_{A}(t^{\prime})\left[aW_{\mathrm{BTZ}}^{M_{1}}(x_{A}(t),x_{A}(t^{\prime}))+bW_{\mathrm{BTZ}}^{M_{2}}(x_{A}(t),x_{A}(t^{\prime}))\right]+(A\leftrightarrow B)\right]\notag\\
		&-|1,1\rangle\langle0,0|(a+b)\left[\int dtdt^{\prime}\bar{\nu}_{A}(t)\bar{\nu}_{B}(t^{\prime})\left[aW_{\mathrm{BTZ}}^{M_{1}}(x_{A}(t),x_{B}(t^{\prime}))+bW_{\mathrm{BTZ}}^{M_{2}}(x_{A}(t),x_{B}(t^{\prime}))\right]+(A\leftrightarrow B)\right].\label{fphi20}
		\end{align}
	where $a=\cos\theta\cos\varphi$, $b=\sin\theta\sin\varphi$.
	Likewise we can also calculate $\mathrm{Tr}_{\phi}[\langle s_{f}|\Phi_{f}^{(0)}\rangle\langle\Phi_{f}^{(2)}|s_{f}\rangle]$.

\section{NUMERICAL EVALUATION OF MATRIX COMPONENT} \label{appa2}
	For the following calculations we focus on two identical static detectors along the same axis at the black hole's center ($\Delta\phi=0$), and assume both detectors switch on and off simultaneously.
	Under these conditions, the expression for the integration of (\ref{c}) is given by
	\begin{align}
		&\int dtdt^{\prime}\nu_A(t)\bar{\nu}_B(t^{\prime})W_{\mathrm{BTZ}}^{M_1M_2}(x_A(t),x_B(t^{\prime}))\notag\\
		=&\frac{\gamma_{\mathrm{A}}^{M_1}\gamma_{\mathrm{B}}^{M_2}}{4\pi l\mathcal{N}\sqrt{2}}\sum_{p,q=-\infty}^{\infty}\int_{\mathbb{R}}dt_{\mathrm{A}}\int_{\mathbb{R}}dt^{\prime}_{\mathrm{B}} e^{-(\gamma_{\mathrm{A}}^{M_1})^2t_{\mathrm{A}}^{2}/2\sigma^{2}}e^{-(\gamma_{\mathrm{B}}^{M_2})^2t_{\mathrm{B}}^{\prime2}/2\sigma^{2}}e^{-i\Omega(\gamma^{M_1}_{\mathrm{A}}t_{\mathrm{A}}-\gamma^{M_2}_{\mathrm{B}}t^{\prime}_{\mathrm{B}})}\left[\frac{1}{\rho^{-}(t_{\mathrm{A}},t^{\prime}_{\mathrm{B}})}-\frac{\zeta}{\rho^{+}(t_{\mathrm{A}},t^{\prime}_{\mathrm{B}})}\right],\label{wab}
		\end{align}
	where $\gamma^{M_i}_D=\sqrt{\frac{r_D^2}{l^2}-M_i}$ is the redshift factor and
		\begin{equation}
		\rho^{\pm}(t_{\mathrm{A}},t^{\prime}_{\mathrm{B}}):=\frac{\sqrt{\gamma_{\mathrm{A}}^{M_1}\gamma_{\mathrm{B}}^{M_2}}}{\sqrt{\sqrt{M_1}\sqrt{M_2}}}\sqrt{\cosh\chi_{\mathrm{AB},pq}^{\pm}-\cosh\left[\frac{\sqrt{M_1}t_A-\sqrt{M_2}t^{\prime}_B}{l}-i\epsilon\right]},\label{rhoab}
		\end{equation}
		\begin{equation}
		\chi_{\mathrm{AB},pq}^{\pm}:=\operatorname{arccosh}\left[\frac{\sqrt{M_1}\sqrt{M_2}}{\gamma_{\mathrm{A}}^{M_1}\gamma_{\mathrm{B}}^{M_2}}\left(\frac{r_{\mathrm{A}}r_{\mathrm{B}}}{l^2\sqrt{M_1}\sqrt{M_2}}\cosh\left[2\pi (p\sqrt{M_1}-q\sqrt{M_2})\right]\pm1\right)\right].\label{arccoshpq}
		\end{equation}
	Further changing coordinates $u:=\sqrt{M_1}t_A-\sqrt{M_2}t^{\prime}_B$, $s:=\sqrt{M_1}t_A+\sqrt{M_2}t^{\prime}_B$, it yields
		\begin{align}
				&\int dtdt^{\prime}\nu_A(t)\bar{\nu}_B(t^{\prime})W_{\mathrm{BTZ}}^{M_1M_2}(x_A(t),x_B(t^{\prime}))\notag\\
				=&\frac{\eta ^{M_1}_{\mathrm{A}}\eta ^{M_2}_{\mathrm{B}}}{8\pi l\mathcal{N}\sqrt{2}}\sum_{p,q=-\infty}^{\infty}\int_{\mathbb{R}}\mathrm{d}ue^{-[(\eta _{\mathrm{A}}^{M_1})^2+(\eta _{\mathrm{B}}^{M_2})^2]u^{2}/8\sigma^{2}}e^{i\Omega(\eta ^{M_1}_{\mathrm{A}}+\eta ^{M_2}_{\mathrm{B}})u/2}\left[\frac{1}{\rho^{-}(u)}-\frac{\zeta}{\rho^{+}(u)}\right]\notag\\
				&\times\int_{\mathbb{R}}\mathrm{d}se^{-[(\eta _{\mathrm{A}}^{M_1})^2+(\eta _{\mathrm{B}}^{M_2})^2]s^{2}/8\sigma^{2}}e^{-[(\eta _{\mathrm{A}}^{M_1})^2-(\eta _{\mathrm{B}}^{M_2})^2]us/4\sigma^{2}}e^{-\mathrm{i}\Omega(\eta ^{M_1}_{\mathrm{A}}-\eta ^{M_2}_{\mathrm{B}})s/2}\notag\\
				=&\frac{\sigma\eta ^{M_1}_\mathrm{A}\eta ^{M_2}_\mathrm{B}}{4l\mathcal{N}\sqrt{\pi((\eta ^{M_1}_\mathrm{A})^2+(\eta ^{M_2}_\mathrm{B})^2)}}\mathrm{exp}\left[-\frac{\Omega^2\sigma^2(\eta ^{M_1}_\mathrm{A}-\eta ^{M_2}_\mathrm{B})^2}{2((\eta ^{M_1}_\mathrm{A})^2+(\eta ^{M_2}_\mathrm{B})^2)}\right]\notag\\
				&\times\sum_{p,q=-\infty}^{\infty}\int_{\mathbb{R}}\mathrm{d}u\exp\left[-\frac{(\eta ^{M_1}_\mathrm{A})^2(\eta ^{M_2}_\mathrm{B})^2u^2}{2\sigma^2((\eta ^{M_1}_\mathrm{A})^2+(\eta ^{M_2}_\mathrm{B})^2)}\right]\exp\left[-\frac{i\eta ^{M_1}_\mathrm{A}\eta ^{M_2}_\mathrm{B}(\eta ^{M_1}_\mathrm{A}+\eta ^{M_2}_\mathrm{B})\Omega u}{(\eta ^{M_1}_\mathrm{A})^2+(\eta ^{M_2}_\mathrm{B})^2}\right]\left[\frac{1}{\rho^-(u)}-\frac{\zeta}{\rho^+(u)}\right]\notag\\
				=&2K^{M_1M_2}\sum_{p,q=-\infty}^{\infty}\mathrm{Re}\int_{0}^{\infty}\mathrm{d}xe^{-\alpha^{M_1M_2} x^{2}}e^{-i\beta^{M_1M_2} x}\left[\frac{1}{\sqrt{\cosh\chi_{\mathrm{AB},pq}^{-}-\cosh x}}-\frac{\zeta}{\sqrt{\cosh\chi_{\mathrm{AB},pq}^{+}-\cosh x}}\right],\label{labm12}
		\end{align}
	where $\eta^{M_i}_D=\frac{\gamma^{M_i}_D}{\sqrt{M_i}}$ and 
		\begin{equation}
			\alpha^{M_1M_2}:=-\frac{(\eta^{M_1}_\mathrm{A})^2(\eta^{M_2}_\mathrm{B})^2l^2}{2\sigma^2((\eta^{M_1}_\mathrm{A})^2+(\eta ^{M_2}_\mathrm{B})^2)},\label{am1m2}
		\end{equation}
		\begin{equation}
		\beta^{M_1M_2}:=\frac{\eta ^{M_1}_\mathrm{A}\eta ^{M_2}_\mathrm{B}(\eta^{M_1}_\mathrm{A}+\eta^{M_2}_\mathrm{B})l\Omega}{(\eta^{M_1}_\mathrm{A})^2+(\eta^{M_2}_\mathrm{B})^2},\label{bm1m2}
		\end{equation}
		\begin{equation}
		K^{M_1M_2}:=\frac{\sigma}{4\mathcal{N}}\sqrt{\frac{\eta^{M_1}_\mathrm{A}\eta^{M_2}_\mathrm{B}}{\pi((\eta^{M_1}_\mathrm{A})^2+(\eta^{M_2}_\mathrm{B})^2)}}\exp\left[-\frac{\Omega^2\sigma^2(\eta^{M_1}_\mathrm{A}-\eta^{M_2}_\mathrm{B})^2}{2((\eta^{M_1}_\mathrm{A})^2+(\eta^{M_2}_\mathrm{B})^2)}\right].\label{km1m1}
		\end{equation}
	In the last step, we defining: $x:=\frac{u}{l}$ enables the final form for numerical evaluation.
	An analogous procedure as in (\ref{labm12}) yields
		\begin{align}
			&\int dtdt^{\prime}\nu_A(t)\bar{\nu}_B(t^{\prime})W_{\mathrm{BTZ}}^{M_i}(x_A(t),x_B(t^{\prime}))\notag\\
			=&2K^{M_i}\sum_{n=-\infty}^{\infty}\mathrm{Re}\int_{0}^{\infty}\mathrm{d}xe^{-\alpha^{M_i} x^{2}}e^{-\mathrm{i}\beta^{M_i} x}\left[\frac{1}{\sqrt{\cosh\chi_{\mathrm{AB,n}}^{M_i-}-\cosh x}}-\frac{\zeta}{\sqrt{\cosh\chi_{\mathrm{AB,n}}^{M_i+}-\cosh x}}\right],\label{labnu}
		\end{align}
	where
		\begin{equation}
			\alpha^{M_i}:=-\frac{(\gamma^{M_i}_\mathrm{A})^2(\gamma^{M_i}_\mathrm{B})^2}{2\sigma^2((\gamma^{M_i}_\mathrm{A})^2+(\gamma^{M_i}_\mathrm{B})^2)}\frac{l^2}{M_i},\label{ami}
		\end{equation}
		\begin{equation}
		\beta^{M_i}:=\frac{\gamma^{M_i}_\mathrm{A}\gamma^{M_i}_\mathrm{B}(\gamma^{M_i}_\mathrm{A}+\gamma^{M_i}_\mathrm{B})}{(\gamma^{M_i}_\mathrm{A})^2+(\gamma^{M_i}_\mathrm{B})^2}\frac{l}{\sqrt{M_i}}\Omega,\label{bmi}
		\end{equation}
		\begin{equation}
		K^{M_i}:=\frac{\sigma}{4}\sqrt{\frac{\gamma^{M_i}_\mathrm{A}\gamma^{M_i}_\mathrm{B}}{\pi((\gamma^{M_i}_\mathrm{A})^2+(\gamma^{M_i}_\mathrm{B})^2)}}\exp\left[-\frac{\Omega^2\sigma^2(\gamma^{M_i}_\mathrm{A}-\gamma^{M_i}_\mathrm{B})^2}{2((\gamma^{M_i}_\mathrm{A})^2+(\gamma^{M_i}_\mathrm{B})^2)}\right],\label{kmi}
		\end{equation}
		\begin{equation}
		\chi_{\mathrm{AB},n}^{M_i\pm}:=\operatorname{arccosh}\left[\frac{M_i}{\gamma_{\mathrm{A}}^{M_i}\gamma_{\mathrm{B}}^{M_i}}\left(\frac{r_{\mathrm{A}}r_{\mathrm{B}}}{l^2M_i}\cosh\left[2\pi n\sqrt{M_i}\right]\pm1\right)\right].\label{arccos}
		\end{equation}
	For (\ref{pdmi}), it is known to be \cite{Henderson:2017yuv}
		\begin{align}
			P_D^{M_i}=&\int dtdt^{\prime}\nu_D(t)\bar{\nu}_D(t^{\prime})W_{\mathrm{BTZ}}^{M_i}(x_D(t),x_D(t^{\prime}))\notag\\
			=&\frac{\sigma^{2}}{2}\int_{\mathbb{R}}\mathrm{d}x\frac{e^{-\sigma^{2}(x-\Omega)^{2}}}{e^{x/T^{M_i}_{\mathrm{D}}}+1}-\zeta\frac{\sigma}{2\sqrt{2\pi}}\mathrm{Re}\int_{0}^{\infty}\mathrm{d}x\frac{e^{-\alpha^{M_i}_{\mathrm{D}}x^{2}}e^{-i\beta^{M_i}_{\mathrm{D}}x}}{\sqrt{\cosh\chi_{\mathrm{D},0}^{M_i+}-\cosh x}}\notag\\
&+\frac{\sigma}{\sqrt{2\pi}}\sum_{n=1}^{\infty}\mathrm{Re}\int_{0}^{\infty}\mathrm{d}xe^{-\alpha^{M_i}_{\mathrm{D}}x^{2}}e^{-i\beta^{M_i}_{\mathrm{D}}x}\left(\frac{1}{\sqrt{\cosh\chi_{\mathrm{D},n}^{M_i-}-\cosh x}}-\frac{\zeta}{\sqrt{\cosh\chi_{\mathrm{D},n}^{M_i+}-\cosh x}}\right),\label{pdminum}
		\end{align}
	where $T^{M_i}_{\mathrm{D}}=\sqrt{M_i}/\left(2\pi l\gamma^{M_i}_{\mathrm{D}}\right)$ is the local temperature at $r=r_\mathrm{D}$ and 
		\begin{align}
		&\alpha^{M_i}_{\mathrm{D}}:=\frac{(\gamma^{M_i}_{\mathrm{D}})^{2}l^{2}}{4\sigma^{2}M_i},\quad\beta^{M_i}_{\mathrm{D}}:=\frac{\gamma^{M_i}_{\mathrm{D}}l\Omega}{\sqrt{M_i}},\label{apdmi}\\
		&\chi_{\mathrm{D},n}^{M_i\pm}:=\operatorname{arccosh}\left[\frac{M_i}{(\gamma^{M_i}_{\mathrm{D}})^{2}}\left(\frac{r_{\mathrm{D}}^{2}}{l^{2}M_i}\cosh\left[2\pi n\sqrt{M_i}\right]\pm1\right)\right].\label{arccoshdmi}
		\end{align}
	The first two terms, corresponding to $\begin{array}{ccc}(n&=&0)\end{array}$, resemble AdS–Rindler contributions in the BTZ  spacetime, whereas the last term $(n\neq0)$ is known as the BTZ term.
	Both the second and third integrals in (\ref{pdminum}) exhibit the same branch cut subtlety as (\ref{labm12}) and (\ref{labnu}), but can be handled in an analogous manner \cite{Bueley:2022ple}.
	
\section{NORMALIZATION OF DENSITY MATRIX} \label{appa3}
The normalized density matrix is obtained by renormalizing $\rho_{AB}$ (\ref{matrix}) through division by its trace norm.
For simplicity, we take $\theta=\varphi$ and the corresponding trace of the unnormalized density matrix is given by [up to $\mathcal{O}(\lambda^{2})$]
		\begin{align}
		\mathrm{Tr}(\rho_{AB})=&P_G+P_A+P_B\notag\\
		=&1-2\lambda^{2}\left[(\tilde{a}^2+\tilde{a}\tilde{b})(P_{A}^{M_{1}}+P_{B}^{M_{1}})+(\tilde{b}^2+\tilde{a}\tilde{b})(P_{A}^{M_{2}}+P_{B}^{M_{2}})\right]\notag\\
		&+\lambda^{2}(\tilde{a}^2P_{A}^{M_{1}}+\tilde{b}^2P_{A}^{M_{2}}+2\tilde{a}\tilde{b}P_{A}^{M_{1}M_{2}})
		+\lambda^{2}(\tilde{a}^2P_{B}^{M_{1}}+\tilde{b}^2P_{B}^{M_{2}}+2\tilde{a}\tilde{b}P_{B}^{M_{1}M_{2}})\notag\\
		=&1-\lambda^{2}\Big[\tilde{a}^2(P_{A}^{M_{1}}+P_{B}^{M_{1}})+\tilde{b}^2(P_{A}^{M_{2}}+P_{B}^{M_{2}})+2\tilde{a}\tilde{b}(P_{A}^{M_{1}}+P_{B}^{M_{1}}+P_{A}^{M_{2}}+P_{B}^{M_{2}}-P_{A}^{M_{1}M_{2}}-P_{B}^{M_{1}M_{2}})\Big],\label{nor}
		\end{align}
	where $\tilde{a}=\cos^{2}\theta$, $\tilde{b}=\sin^{2}\theta$ [by putting $\theta=\varphi$ in (\ref{ab})].
	The entries of the normalized density matrix $\tilde{\rho}_{AB}$ up to order $\lambda^{2}$ are now given by
		\begin{align}
		\tilde{P}_{G}=&P_{G}\mathrm{Tr}(\rho_{AB})^{-1}\notag\\
		=&\frac{1-2\lambda^{2}\left[(\tilde{a}^2+\tilde{a}\tilde{b})(P_{A}^{M_{1}}+P_{B}^{M_{1}})+(\tilde{b}^2+\tilde{a}\tilde{b})(P_{A}^{M_{2}}+P_{B}^{M_{2}})\right]}{1-\lambda^{2}\left[\tilde{a}^2(P_{A}^{M_{1}}+P_{B}^{M_{1}})+\tilde{b}^2(P_{A}^{M_{2}}+P_{B}^{M_{2}})+2\tilde{a}\tilde{b}(P_{A}^{M_{1}}+P_{B}^{M_{1}}+P_{A}^{M_{2}}+P_{B}^{M_{2}}-P_{A}^{M_{1}M_{2}}-P_{B}^{M_{1}M_{2}})\right]}\notag\\
		\simeq&\Bigg[1-2\lambda^{2}\left[(\tilde{a}^2+\tilde{a}\tilde{b})(P_{A}^{M_{1}}+P_{B}^{M_{1}})+(\tilde{b}^2+\tilde{a}\tilde{b})(P_{A}^{M_{2}}+P_{B}^{M_{2}})\right]\Bigg]\notag\\
		&\times\Bigg[1+\lambda^{2}\left[\tilde{a}^2(P_{A}^{M_{1}}+P_{B}^{M_{1}})+\tilde{b}^2(P_{A}^{M_{2}}+P_{B}^{M_{2}})+2\tilde{a}\tilde{b}(P_{A}^{M_{1}}+P_{B}^{M_{1}}+P_{A}^{M_{2}}+P_{B}^{M_{2}}-P_{A}^{M_{1}M_{2}}-P_{B}^{M_{1}M_{2}})\right]\Bigg]\notag\\
		=&1+\lambda^{2}\left[\tilde{a}^2(P_{A}^{M_{1}}+P_{B}^{M_{1}})+\tilde{b}^2(P_{A}^{M_{2}}+P_{B}^{M_{2}})+2\tilde{a}\tilde{b}(P_{A}^{M_{1}}+P_{B}^{M_{1}}+P_{A}^{M_{2}}+P_{B}^{M_{2}}-P_{A}^{M_{1}M_{2}}-P_{B}^{M_{1}M_{2}})\right]\notag\\
		&-2\lambda^{2}\left[(\tilde{a}^2+\tilde{a}\tilde{b})(P_{A}^{M_{1}}+P_{B}^{M_{1}})+(\tilde{b}^2+\tilde{a}\tilde{b})(P_{A}^{M_{2}}+P_{B}^{M_{2}})\right]-\mathcal{O}(\lambda^{4})\notag\\
		=&1-\lambda^{2}(\tilde{a}^2P_{A}^{M_{1}}+\tilde{b}^2P_{A}^{M_{2}}+2\tilde{a}\tilde{b}P_{A}^{M_{1}M_{2}})-\lambda^{2}(\tilde{a}^2P_{B}^{M_{1}}+\tilde{b}^2P_{B}^{M_{2}}+2\tilde{a}\tilde{b}P_{B}^{M_{1}M_{2}})-\mathcal{O}(\lambda^{4})\notag\\
		=&1-P_A-P_B-\mathcal{O}(\lambda^{4}),\label{norpg}
		\end{align}
		\begin{equation}
		\tilde{P}_{D}=P_{D}\mathrm{Tr}(\rho_{AB})^{-1}=P_{D}+\mathcal{O}(\lambda^{4}),\quad\tilde{\mathcal{L}}_{\mathrm{AB}}=\mathcal{L}_{\mathrm{AB}}\mathrm{Tr}(\rho_{AB})^{-1}=\mathcal{L}_{\mathrm{AB}}+\mathcal{O}(\lambda^{4}),\quad\tilde{\mathcal{M}}=\mathcal{M}\mathrm{Tr}(\rho_{AB})^{-1}=\mathcal{M}+\mathcal{O}(\lambda^{4}).
		\end{equation}
\end{widetext}

%
\end{document}